\documentclass[pre]{revtex4}

 \usepackage{amsmath}
  \usepackage{paralist}
  \usepackage{graphics} 
  \usepackage{epsfig} 
 \usepackage[colorlinks=true]{hyperref}

\hypersetup{urlcolor=blue, citecolor=red}

\usepackage{amssymb}
\usepackage{mathrsfs}
\usepackage{bm}
\usepackage{rotating}
\usepackage{cancel}
\usepackage{wasysym}
\usepackage[normalem]{ulem}
\usepackage{multirow}

\begin{document}

\preprint{APS/123-QED}

\title{On the Generalized Hermite-Based Lattice Boltzmann Construction, Lattice Sets, Weights, Moments, Distribution Functions   and High-Order Models}

\author{Ra\'ul Machado}
 \email{raul\_machado@hotmail.com} 
\affiliation{%
Faculty of Engineering and the Environment, University of Southampton, Southampton, SO17 1BJ, United Kingdom}%


\begin{abstract}
The influence of the use of the generalized Hermite polynomial on the Hermite-based lattice Boltzmann (LB) construction approach, lattice sets, the thermal weights, moments and the equilibrium distribution function (EDF) are addressed. 
A new moment \textit{system} is proposed. The theoretical possibility to obtain a high-order Hermite-based LB model capable to exactly  match some first hydrodynamic moments thermally 1) \textit{on}-Cartesian lattice, 2) with \textit{thermal} weights in the EDF, 3) whilst the highest possible hydrodynamic moments that are exactly matched are obtained with the \textit{shortest} \textit{on}-Cartesian lattice sets with some fixed real-valued temperatures, is also analyzed. 
\end{abstract}


\maketitle

\section{Introduction}\label{section:IntroductionHigherOrder}

\begin{figure}
\centering
	\begin{center}
\epsfig{file=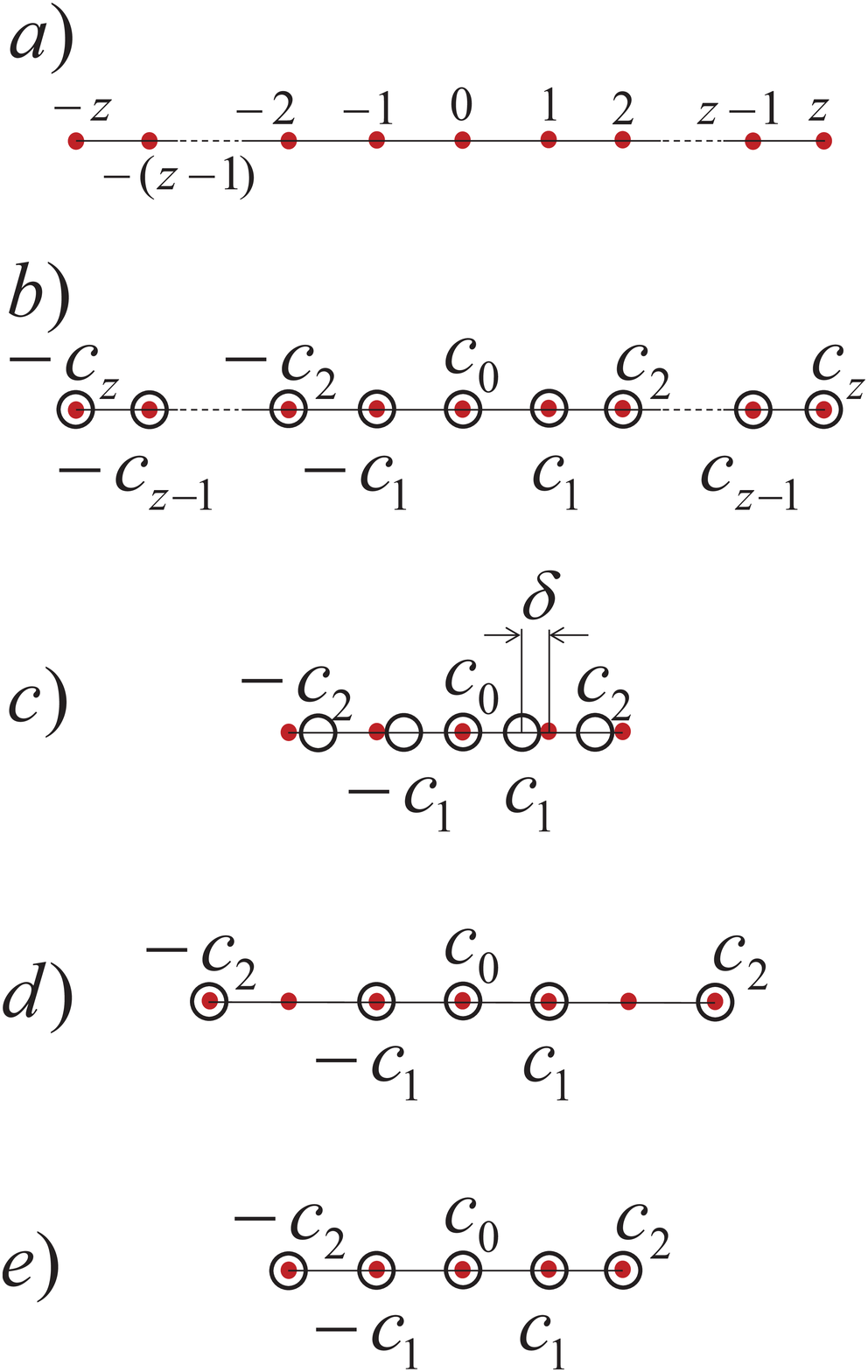,angle=0,width=0.40 \linewidth,clip=}  
	\caption{\label{fig:Cartesian}  (Color online) Schematic representation of the relationship between the Cartesian grid and the discrete lattice velocities. Symbols: Dot (\textcolor{red}{\scriptsize{$\CIRCLE$}}): Nodes in the Cartesian grid; Circle (\textcolor{black}{\Huge{$\circ$}}): Discrete lattice velocities $c_i$. 
	a) One-dimensional Cartesian grid. 
	b) One-dimensional \textit{on}-Cartesian lattice, where all $c_i=i$, $i = $ integer. 
	c) One-dimensional \textit{off}-Cartesian lattice, where there exist at least some $c_i \neq i$, $i = $ integer and $\delta$ is one of the Cartesian-lattice mismatch distance. 
	d) One-dimensional \textit{on}-Cartesian lattice set $\textbf{c}= \{ 0, \pm 1, \pm 3 \} $, i.e. it is \textit{not} the shortest lattice set for D1Q5. 
	e) One-dimensional \textit{on}-Cartesian lattice set $\textbf{c}= \{ 0, \pm 1, \pm 2 \} $, i.e. it is \textit{indeed} the shortest lattice set for D1Q5. 
  One-dimensional low-order LB models have $z=1$, while one-dimensional high-order LB models have $z >1$. }
	\end{center}
\end{figure}

The lattice Boltzmann (LB) method has been used as a viable alternative for numerical simulation of (isothermal) fluid flows  for more than two decades \cite{McNamaraZanetti}, \cite{HigueraJimenez1989}, \cite{HigueraSucciBenzi1989}, \cite{Koelman1991}, \cite{ChenChenMatthaeus1992}, \cite{QianHumiereLallemand1992}, \cite{SucciBook}, \cite{HanelGermanBook2004}. Yet, many aspects regarding the LB method can be debated, such as the choice of the construction approach to build the LB model. The continuous Boltzmann equation can be particularly discretized in both time and phase space \cite{HeLuoDerivationLB}, leading to the LB equation 
\begin{eqnarray} \label{eq:LB}
f_i(\boldsymbol{x}+\boldsymbol{c}_i \delta t,t+\delta t) = f_i(\boldsymbol{x},t) + \mathcal{Q}(f_i).
\end{eqnarray}
Eq. \eqref{eq:LB} is a discrete kinetic equation for populations $f_i(\boldsymbol{x},t)$, where $i=1, \dots, n_q$ and $n_q$ is the number of discrete lattice velocity vectors on a Cartesian grid.  $f_i$ 
 represents the probability of finding a particle with velocity $\boldsymbol{c}_i$ at position $\boldsymbol{x}$ and time $t$ in lattice units \cite{SucciBook}.  
$\mathcal{Q}(f_i)$ is the collision vector. The insertion of the nonlinear Bhatnagar-Gross-Krook (BGK)  \cite{BhatnagarGrossKrook} or Welander \cite{Welander} collision model $\mathcal{Q}(f_i) = -1/\tau (f_i - f_i^{\textrm{eq}})$ into Eq. \eqref{eq:LB} leads to the LBGK equation. 
Note that the $f_i$ inside $\mathcal{Q}(f_i)$ is computed at time $t$, i.e. the LB method is explicit. 
$\tau$ is the relaxation time, non-dimensionalized with $\delta t$, which describes the time of a perturbed system to return to equilibrium and it is related to the viscosity of the fluid. LB models are usually denoted as D$d$Q$n_q$, where $d$ is the dimension of the model \cite{QianHumiereLallemand1992}. For the one-dimensional ($d=1$) case, LB models with a lattice set of $z=1$ (c.f. Fig. \ref{fig:Cartesian}) are low-order, while those with $z>1$ are high-order (more about this below). The space dependence is dealt by summing over all the nodes of the lattice. For instance, for a low-order one-dimensional LB model, the $n_q=3$ and thus $\sum_{i=0}^{n_q-1} f_i c_i^M = - c_1^M f_2 + c_0^M f_0  + c_1^M f_1$, where $c_0=0$ and $M$ is a non-negative integer (more about this below). These LB constructions with integer $c_i$ values are denoted as \textit{on}-Cartesian lattice models, while those with any non-integer $c_i$ value are called \textit{off}-Cartesian lattice (Fig. \ref{fig:Cartesian}).   
The importance of the LB equations, the asymptotic convergence to the continuum Boltzmann-BGK equation, the comparison to the Grad 13 moment system, etc are summarized in \cite{Shan2011}.  



Generally, LB modeling boils down to find an equilibrium distribution function (EDF), $f_i^{\textrm{eq}}$, so that some hydrodynamic moments, e.g. Maxwell-Boltzmann (MB) (convective) $M$-moments 
\begin{eqnarray} \label{eq:MB-convective-moments}
\sum_{i=0}^{n_q-1} f_i^{\textrm{eq}} \boldsymbol{c}^{(M)} =   \rho \ \textrm{\Large{\textit{e}}}\Big(-\frac{\boldsymbol{u}^2}{2 \theta}\Big) \theta^M 
\frac{\partial^M }{\partial \boldsymbol{u}^M} \Bigg(  \textrm{\Large{\textit{e}}}\Big(\frac{\boldsymbol{u}^2}{2 \theta}\Big)  \Bigg),  
\end{eqnarray}
are matched. $\boldsymbol{c}^{(M)}=\boldsymbol{c} \cdots \boldsymbol{c}$, $M$-times and $\textrm{\Large{\textit{e}}}(x)$ is the  classical exponential function. $\rho$ is the density, $\boldsymbol{u}$ is the flow velocity and $\theta=R T$, where $R$ is the specific gas constant and $T$ is the temperature. The right hand side of Eq. \eqref{eq:MB-convective-moments} are the MB moments from which the density, momentum density, pressure tensor, energy flux, rate of change of the energy flux conservations  are obtained with $M=0, 1, 2, 3, 4$ respectively. Eq. \eqref {eq:MB-convective-moments} is well known from the literature, c.f. Eq. (20) in \cite{ChenShanPhysicaD2008}, and its equivalent, Eqs. (14) and (5) in \cite{MachadoRaulSurfaceReaction2012} and \cite{MachadoMATCOMBC} respectively. 
The link between the needed lattice velocities $n_q$ to match high order moments \eqref{eq:MB-convective-moments} is discussed below. 

From now on, the words thermal and isothermal are usually stated in this work in conjunction with the hydrodynamic moments, e.g. r.h.s. of Eq. \eqref{eq:MB-convective-moments}.  By thermal means that $\theta$ does not need to be equal to $\theta_0$ in order to match (some) hydrodynamic moments, where $\theta_0$ is a \textit{particular} fixed value needed to match certain hydrodynamic moment. $\theta_0 = R T_0$. In this context, isothermal means that $\theta=\theta_0$. How this particular $\theta_0$ should be, is presented below (e.g. in connection with tables \ref{tabular:ReferenceTemperatureD1Qnq} and  \ref{tabular:GeneralizedReferenceTemperatureD1Qnq}). Sometimes, $\theta_0$ is denoted as reference ``temperature''. Similarly, thermal and athermal (or isothermal) weights are denoted to those weights that are $\theta$-dependent and $\theta_0$-dependent respectively.  

The essence of the main LB idea is captured by Sauro Succi in \cite{SucciBook},\cite{SucciLecture2006} 
and strengthen in \cite{BrownleeGorbanLevesley2008} with the statement: ``Nonlinearity is local, non-locality is (a) linear; (b) exact and explicitly solvable for all time steps; (c) space discretization is an exact operation''. 
Furthermore, those theoretically fulfilled conservation laws, e.g. \eqref{eq:MB-convective-moments}, (depending on the chosen LB model) are mathematically matched exactly and computationally matched to machine roundoff. To the LBM assets can be added: inherently parallelizable, easy handling on geometries located \textit{on}-Cartesian lattice, free of interpolations, finite difference schemes and correcting (counter) terms (i.e. with no added extra terms evaluated using finite-difference schemes to obtain certain desired property).

The low-order LBGK models contain lattices suitable to reconstruct the Navier-Stokes equation close to the incompressible limit \cite{QianHumiereLallemand1992}, \cite{HeLuoTheory1997}, \cite{ShanHeDiscretization1998}, \cite{ChenShanPhysicaD2008}. These models fulfill the relation \eqref{eq:MB-convective-moments} up to $M=1$ or 2, and are usually isothermals (i.e. $\theta=\theta_0$), when they are free of correcting counter terms. 
A more free $\theta$ value can be theoretically obtained in these models at the expense of the existence of spurious velocity terms. These in turn can be corrected/annihilated by adding extra terms evaluated using finite-difference scheme (i.e. correcting counter terms). However, such approach does not guarantee the main LB idea.


High-order lattice  are also studied in the literature, c.f. \cite{PhilippiHegeleSantosSurmas2006}, \cite{ShanYuanChen2006}, \cite{ShanChen2007}, \cite{SiebertHegelePhilippiIJMPC2007}, \cite{ChenShanPhysicaD2008}, \cite{NieShanChen2008}, \cite{KimPitschBoyd2008}, \cite{TangZhangEmerson2008}, \cite{Shan2010}, \cite{MengZhangPRE2011}, \cite{ChikatamarlaKarlin1D2006}, \cite{KarlinChikatamarla2008CPC}, \cite{ChikatamarlaKarlin2009}, where some of those models are (claiming to be) capable to recover hydrodynamics beyond the Navier-Stokes equation. 
These constructions match the expression \eqref{eq:MB-convective-moments} up to $M$-moments,  $M>2$, with a certain degree of accuracy. The last three aforecited works, \cite{ChikatamarlaKarlin1D2006}, \cite{KarlinChikatamarla2008CPC}, \cite{ChikatamarlaKarlin2009}, are based on the so called ``entropic'' lattice Boltzmann (ELB) approach (c.f. appendix), while the rest are what can be called  Hermite-based constructions. 
These high-order ELB models are \textit{on}-Cartesian lattice but isothermals LB constructions with isothermal weights and spurious velocity terms, c.f. \cite{ChikatamarlaKarlin2009} (more about this below). 

Some  characteristics are now outlined for the aforecited Hermite-based LB models: In \cite{PhilippiHegeleSantosSurmas2006}, \cite{SiebertHegelePhilippiIJMPC2007}, two-dimensional thermal models are described with discrete velocity sets but with  athermal weights, c.f. tables 1 and 2 in \cite{SiebertHegelePhilippiIJMPC2007}; A kinetic theory study is address in \cite{ShanYuanChen2006}, where off-Cartesian lattice sets and athermal weights are outlined in tables  1, 2, 3 therein.  A LB model with multiple relaxation time is found in \cite{ShanChen2007}, where the numerical verification is based on off-Cartesian lattice sets and athermal weights, c.f. table 1 therein;  
The accuracy of the (thermal) lattice Boltzmann is studied in \cite{ChenShanPhysicaD2008}; 
A multiple relaxation time LB model is also described in \cite{NieShanChen2008}, where three-dimensional numerical validations are carried out using off-Cartesian with athermal weights, c.f. table 1 therein;
A finite difference scheme is employed in \cite{KimPitschBoyd2008} in an isothermal LB model with off-Cartesian lattice with athermal weights, c.f. tables 1,2 therein. In \cite{Shan2010}, an on-Cartesian Hermite-based LB model is presented, but still it is based on athermal weights and a general construction to obtain the shortest lattice sets are not found in the literature (more about this below).  Because of possible discontinuities at the wall, a finite difference method is chosen in  \cite{MengZhangPRE2011}, due to the presence of off-Cartesian lattice construction. In general, finite difference schemes are adopted in many high-order LB models for stability issues \cite{McNamaraGarciaAlder1995}. 

There exist some other alternative (high-order) LB constructions, e.g. \cite{CaoChenJinMartinez},  \cite{KataokaTsutahara2004}, \cite{QuShuChew2007} and subsequent works, c.f. \cite{ChenXuZhangLiSucci}, \cite{NejatAbdollahi2013}. Unfortunately, finite difference schemes are required.  Hybrid LB constructions can be added to this group. For instance, an LB model is proposed in \cite{LallemandLuo2003}, where mass and conservation equations are solved due to \cite{Humiere1992}, whereas the diffusion-advection equation for the temperature is solved separately, e.g. by using finite-difference. 

It is useful to have high-order thermal LB models on-Cartesian lattice, with thermal weights (based on the final results that are used in the EDF), and with the shortest lattice sets when possible.  Locality has been long recognized as an important source of efficiency in parallel computing to lower communications overhead, c.f. \cite{XuLuBook}. Therefore, for the sake of (parallel) computational cost, it is good to have high-order LB models with consecutive lattice sets, e.g. in one-dimension (Fig. \ref{fig:Cartesian}) $c_i$ = consecutive integers up to z, and thus with the shortest lattice sets. A computational cheap LB construction makes feasible to have a complete (i.e. non-reduced), or at least a less reduced lattice set, needed to match (some) hydrodynamic moments. The importance of weights becomes clear at walls, where the EDF is (almost) equal to the density-scaled weights, $f_i^{\textrm{eq}}=\rho W_i(1+C)$, where $C = $ function($\theta, u$). This, due to the flow velocity is (almost) zero at the walls, depending of the regime (e.g. slip or non-slip flow) and $C = 0$ when $u=0$, regardless $\theta$.  Hence, the importance of having thermal weights is evidenced for walls with $\theta \neq \theta_0$. 

Strategies, such as (but not limited to) interpolations and/or approximations, are sometimes implemented to deal with this  Cartesian-lattice mismatch (c.f. Fig. \ref{fig:Cartesian} c) ). It should be pointed out that with the use of interpolations, the exact matching of the conservation laws is not guarantee and/or the locality is lost for many existing schemes \cite{SucciBook}, \cite{Chenetal2006}. All of this to the detriment of the main LB idea. 
An example: In \cite{TangZhangEmerson2008}, the off-Cartesian lattice problem (Fig. \ref{fig:Cartesian}) is tackled so that the pointwise interpolations are avoided by adopting approximations of the non-integer values to an appropriate (closest) lattice grid point. Their D2Q13 
model with athermal weights is already (claiming to be) able to capture some of the microflows features, although it is recognized in \cite{TangZhangEmerson2008} that a higher order LB model is definitely needed (to match higher order moments and to improve accuracy). Their experience uncovers that moving from standard D2Q9 to their approximated D2Q13 implies not much difference in the computational cost and instability, \cite{TangZhangEmerson2008-PC}. However, the approximation implemented in D2Q13 becomes difficult for D2Q16 and interpolations are needed and thus, the computational cost increases significantly. The D2Q21 was tried,  \cite{TangZhangEmerson2008-PC}, with increased computational cost and serious instabilities despite additional interpolations.

Because of existence of interpolations or approximations (e.g. in some previous Hermite-based LB due to \textit{off}-Cartesian lattice models), spurious velocity terms (e.g. in ELB method due to its macroscopic description property, c.f. appendix) and finite difference schemes (e.g. in alternative LB models), the main LB idea is compromised in some of the aforementioned \textit{high-order} LB models. 


In general, two main issues are addressed in this work: $i$) The influence of the use of the generalized Hermite polynomial on the Hermite-based LB construction approach, lattice sets, the thermal weights, moments and  the equilibrium distribution function. A new moment \textit{system} is proposed. This is handled in sections \ref{section:GeneralizedHermite} and \ref{section:OnTheHigherOrder}. $ii$) An answer is given to the following question: 
Is it (theoretically) possible to obtain a one-dimensional high-order Hermite-based LB model capable to exactly match the first hydrodynamic $z$-moments thermally 1) \textit{on}-Cartesian lattice, 2) with thermal weights (based on the final results that are used in the EDF), 3) whilst the hydrodynamic $(z+1)$-moments are exactly matched with the shortest \textit{on}-Cartesian lattice sets with some fixed real-valued $\theta$? 
This is handled in section \ref{section:OnTheHigherOrder}. 

This is a theoretical work, where the necessary equations are presented in a compact yet complete form, in order to avoid bulky relations. Numerical studies are presented elsewhere. The approach of presenting solely theoretical results about LB prior numerical simulations is adopted by other authors as well, c.f. \cite{QianHumiereLallemand1992}, \cite{ShanHeDiscretization1998}, \cite{KarlinFerrante1999}, \cite{PhilippiHegeleSantosSurmas2006}, \cite{ChenGoldhirschOrszag2008}, \cite{RubinsteinLuo2008}, \cite{Shan2010}. 

\section{On the Generalized Hermite-Based Lattice Boltzmann Construction}\label{section:GeneralizedHermite}

\begin{figure}
\centering
	\begin{center}
\begin{tabular}{cc}
a)  \ \epsfig{file=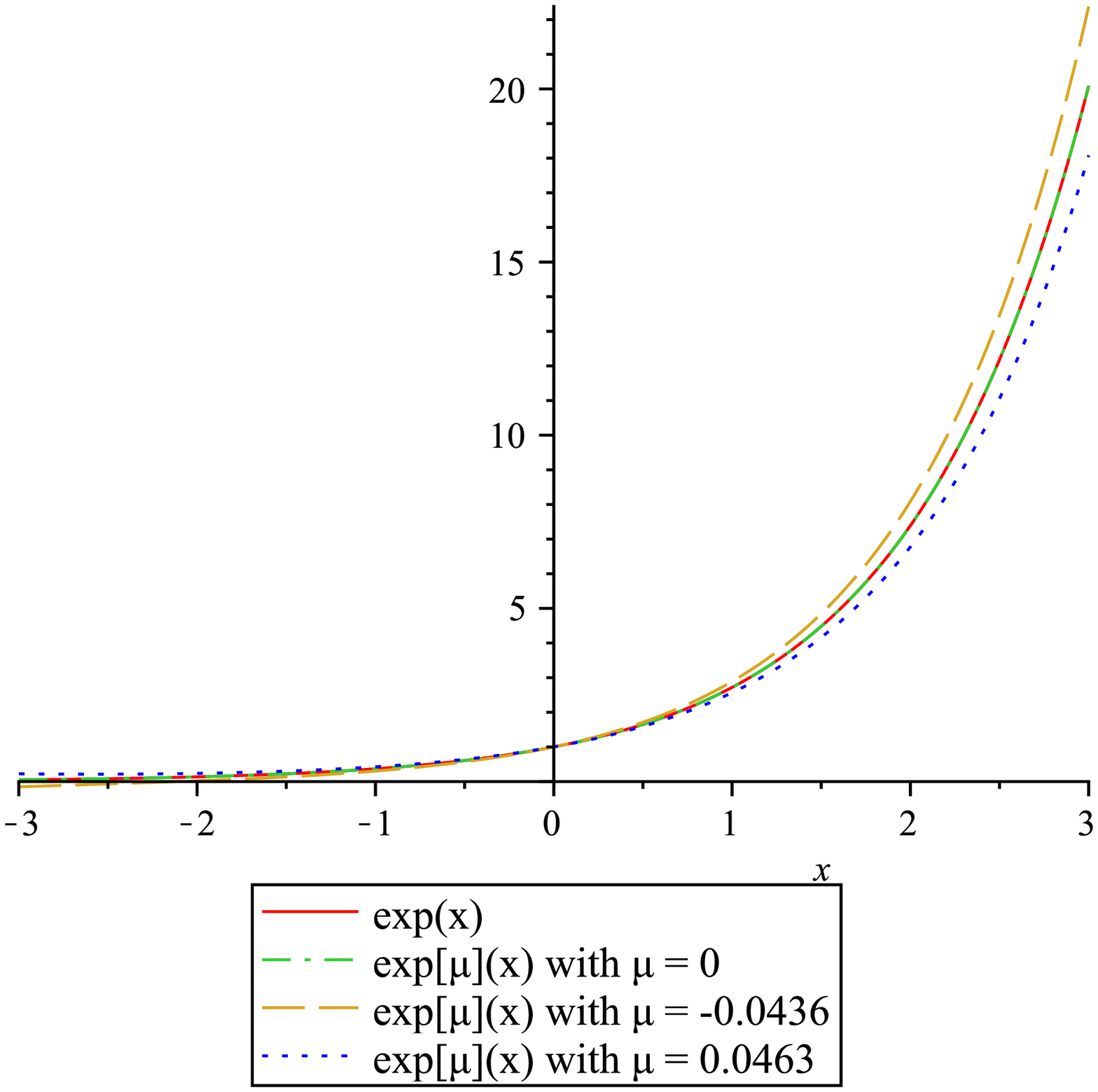,angle=0,width=0.44 \linewidth,clip=}  
b)  \epsfig{file=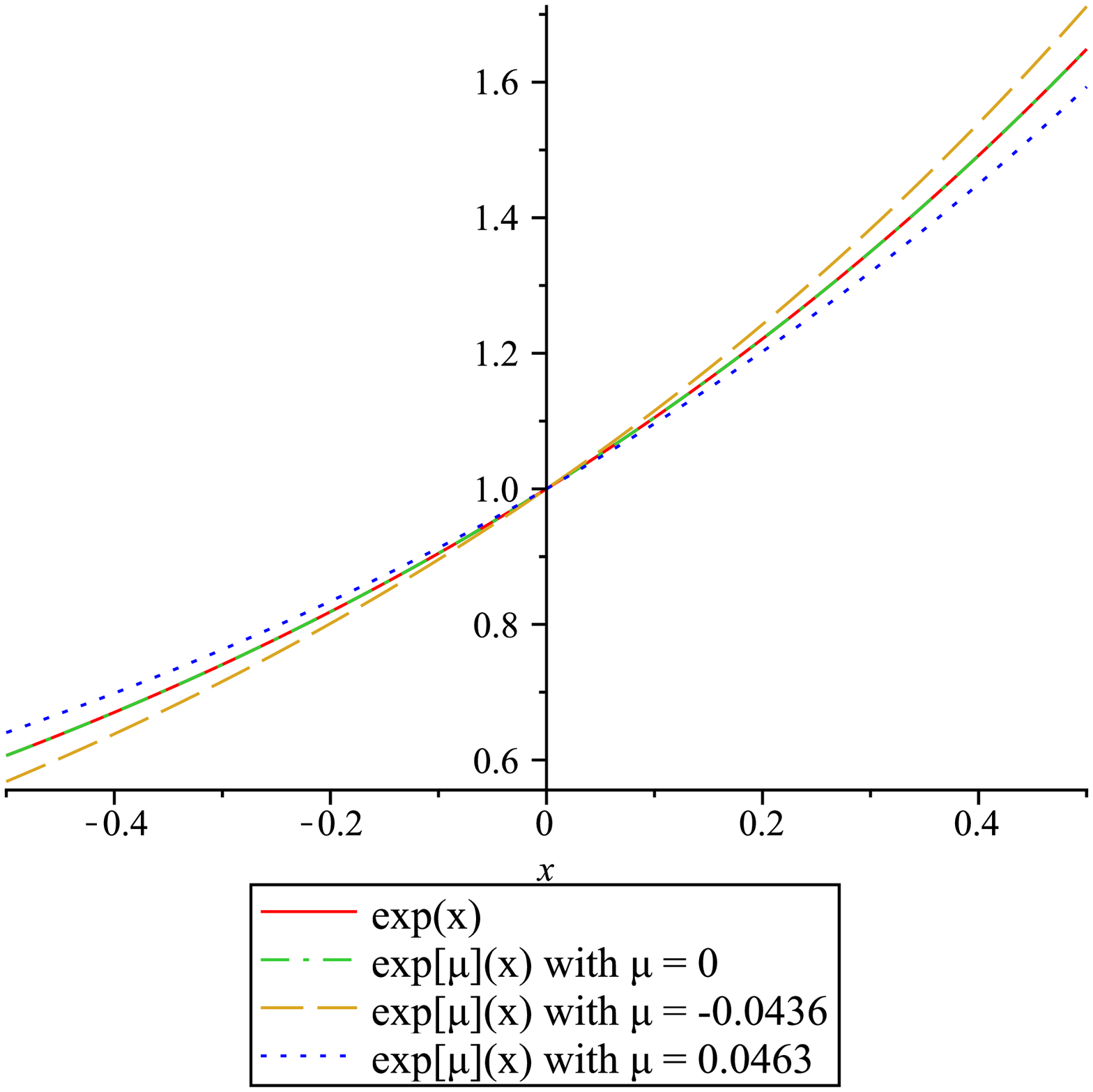,angle=0,width=0.44 \linewidth,clip=} 
\end{tabular}
	\caption{\label{fig:ComparisonExpWithWhitakker}  (Color online) Comparison between the classical exponential $\textrm{\Large{\textit{e}}}(x)$ and the generalized exponential function \eqref{eq:GeneralizedExponentialFunction} with $\mu = -\frac{1}{2} \frac{3 \theta - 1}{\theta}$. a): $\theta = 1/3 \pm 10^{-2}$; b): Zoomed part of a).}
	\end{center}
\end{figure}

It is always recommended to deal with a general formulation when relations are derived. From the classical MB moments \eqref{eq:MB-convective-moments}, the classical exponential function  $\textrm{\Large{\textit{e}}}(x)$ is noticed. 
This suggest its extensions to the term $\textrm{\Large{\textit{e}}}_{\mu}(x)$, which is the generalized exponential function, and it is defined as  
\begin{eqnarray} \label{eq:GeneralizedExponentialFunction}
	\textrm{\Large{\textit{e}}}_{\mu}(x) = (2 x)^{-1/2 - \mu} \  \textrm{WM}_{-1/2,\mu} (2 x), 
\end{eqnarray}
where WM is the Whittaker M-function \cite{BatemanVolI-II-III}, defined as 
\begin{eqnarray} \label{eq:GWhittakerM-function}
\textrm{WM}_{-1/2,\mu}(2 x)=2^{2 \mu} x J_{(-1/2+ \mu)}(x) \Gamma(\frac{1}{2}+\mu)+2^{2 \mu} x J_{(1/2+\mu)}(x) \Gamma(\frac{1}{2}+\mu),
\end{eqnarray}
$\Gamma(n)=(n-1)!$ and $J_{\varsigma}(x)$ is the Bessel function of the first kind, i.e.
\begin{eqnarray} \nonumber
J_{\varsigma}(x) = \sum_{m=0}^{\infty} \frac{(-1)^m}{m! \, \Gamma(m+\varsigma+1)} {\left(\tfrac{1}{2}x\right)}^{2m+\varsigma}.
\end{eqnarray}
The difference between the classical exponential, $\textrm{\Large{\textit{e}}}(x)$, and its generalization, $\textrm{\Large{\textit{e}}}_{\mu}(x)$, is visualized in Fig. \ref{fig:ComparisonExpWithWhitakker} for some $\mu$ values. 
The $\textrm{\Large{\textit{e}}}_{\mu}(x) < \textrm{\Large{\textit{e}}}(x)$ for $x>0$ with $\mu>0$ and for $x<0$ with $\mu<0$.  The opposite, $\textrm{\Large{\textit{e}}}_{\mu}(x) > \textrm{\Large{\textit{e}}}(x)$, is obtained for $x<0$ with $\mu>0$ and for $x>0$ with $\mu<0$. The $\textrm{\Large{\textit{e}}}_{\mu}(x)$ and thereby the $\textrm{\Large{\textit{e}}}_{\mu}(x)$-dependent moment system are reduced to their classical $\textrm{\Large{\textit{e}}}(x)$
and MB moment system \eqref{eq:MB-convective-moments} respectively when $\mu=0$. 
In this context, the generating function for the generalized Hermite polynomial $H_n^{(\mu)}(x)$ is \cite{BatemanVolI-II-III}
\begin{eqnarray} \label{eq:GeneratingFunctionGeneralizedHermitePolynomial}
	\textrm{\Large{\textit{e}}}_{\mu}(2xa) \textrm{\Large{\textit{e}}}(-a^2) = \sum_{n=0}^{\infty} H_n^{(\mu)}(x) \frac{a^n}{n!}. 
\end{eqnarray}
The generalized Hermite polynomials $H_n^{(\mu)}(x)$, introduced by G\'abor Szeg\H{o} \cite{SzegoGHP1939}, is obtained from the   relations  
\begin{subequations} \label{eq:Szego}
\begin{eqnarray} \label{eq1:Szego}
	H_{2n}^{(\mu)}(x) &=& (-1)^n 2^{2n} n! L_n^{\mu-1/2}(x^2), \\ \label{eq2:Szego}
	H_{2n+1}^{(\mu)}(x) &=& (-1)^n 2^{2n+1} n! x L_n^{\mu+1/2}(x^2), 
\end{eqnarray}
\end{subequations}
where $\mu > -1/2$ and $L_n^{\alpha}(x)$ is the generalized Laguerre polynomials \cite{BatemanVolI-II-III}. However, the polynomials obtained from  \eqref{eq:Szego} are sometimes normalized, c.f. \cite{ChiharaPhDThesis1955}, \cite{DickinsonWarsi1963}, \cite{DuttaChatterjeaMore1975}, \cite{RosenblumMarvin1993}. The implemented normalization in this work is 
\begin{eqnarray} \label{eq:Normalization}
	\mathcal{N}_n(a)= \frac{B(\mu, n + a)}{B(\mu, 1/2)}, 
\end{eqnarray}
where $B(x,y)=\Gamma(x)\Gamma(y)/\Gamma(x+y)$ is the beta function and $\Gamma(x)=(x-1)!$ is the gamma function so that 
\begin{subequations} \label{eq:NormalizedSzego}
\begin{eqnarray} \label{eq1:NormalizedSzego}
	H_{2n}^{(\mu)}(x) &=& \mathcal{N}_{n}(1/2) \ H_{2n}^{(\mu)}(x), \\ \label{eq2:NormalizedSzego}
	H_{2n+1}^{(\mu)}(x) &=& \mathcal{N}_{n}(3/2) \ H_{2n+1}^{(\mu)}(x), 
\end{eqnarray}
\end{subequations}
for $n>0$ while $H_{0}^{(\mu)}(x)=1$. The generalized Hermite polynomials used in this work are calculated from Eqs. \eqref{eq:NormalizedSzego}.

\subsection{The Thermal Weights}\label{section:Weights}

The generalized Hermite-based LB construction approach is proposed in this work. Based on the definition of the generalized Hermite polynomial, the LB construction is not valid for $\mu = 1/2 - n$, $n=1,  2, 3, \dots$, which will be seen when an EDF example is presented. The thermal weights are acquired so that they and the abscissas form a generalized Hermite quadrature. The $d$-dimensional weights for the LB D$d$Q$n_q$ models are obtained from 
\begin{eqnarray} \label{eq:muGeneralizedHermiteQuadrature}
	\sum_{i=0}^{n_q - 1} W_i \prod_{\alpha}^d H_n^{(\mu)}(a) = A,
\end{eqnarray}
where $a=c_{\alpha, i}/\sqrt{2 \theta}$, $A=1$ for $\prod_{\alpha}^d H_0^{(\mu)}(a)$, i.e. generalized Hermite order $n=0$, or $A=0$ otherwise and $\alpha=\{ x$, y, z \} in \eqref{eq:muGeneralizedHermiteQuadrature}. $n_q$ is the number of discrete  lattice velocity vectors. A number of $n_q +1$ relations are obtained from \eqref{eq:muGeneralizedHermiteQuadrature}, and the generalized Hermite order $n$ goes from zero to $n_q$.  
For simplicity, this work is focused to a \textit{one-dimensional} ($d=1$) study from now on, i.e. D1Q$n_q$. 
However, this is not a limitation. Two- and three-dimensional weights can be obtained from algebraic products of the one-dimensional weights; for instance, it is well known that the athermal weights $W_0^{*}=2/3$ and $W_{1,2}^{*}=1/6$ from the one-dimensional low-order LB models can be used to construct the two dimensional weights $W_0 = W_0^{*} \cdot W_0^{*}=4/9$, $W_{1-4} = W_0^{*} \cdot W_1^{*}=1/9$ and $W_{5-8} = W_1^{*} \cdot W_2^{*}=1/36$ \cite{QianHumiereLallemand1992}. The same procedure applies for the thermal weights obtained from the formulation \eqref{eq:muGeneralizedHermiteQuadrature} corresponding to the low- and high-order LB models, c.f. \cite{MachadoMATCOMBC}. 
The result is that $\alpha=x$ now and terms such as $c_{\alpha, i}$ and $u_{\alpha}$  are equivalent to $c_{x, i}$ and $u_{x}$ or just simply to $c_{i}$ and $u$. (Do not mix the $z$ parameter seen in Fig. \ref{fig:Cartesian}, with the axis coordinate z, which is no longer used in this work).
The term 
\begin{eqnarray} \label{eq:z-nq-definition}
	z=\frac{n_q-1}{2} 
\end{eqnarray}
is now used throughout this work. The discrete lattice velocities are contained within the vector $\boldsymbol{c} = \{  -c_z, \dots -c_1, 0, c_1, \dots, c_z \} $  for a $d=1$ case, c.f. Fig. \ref{fig:Cartesian}.


The results from Eq. \eqref{eq:muGeneralizedHermiteQuadrature} for the D$1$Q$n_q$ generalized Hermite-based LB model can be formulated as 
\begin{subequations} \label{eq:1DmuGeneralizedWeights}
\begin{eqnarray} \label{eq:1DmuGeneralizedWeights0}
	W_0 &=& \prod_{n=1}^{z} \Bigg( 1- \frac{B \ \theta}{c_n^2}  \Bigg),  \\ \label{eq:1DmuGeneralizedWeightsk}
	W_{\pm c_k} &=& \frac{1}{2} \frac{\theta (2 \mu + 1) }{c_k^2} \prod_{n=1, n \neq k}^{z} \Bigg( 1- \frac{A \ \theta - c_k^2}{c_n^2 - c_k^2}  \Bigg),
\end{eqnarray}
\end{subequations}
where the Pochhammer symbol $(\mu + \frac{1}{2} + a)_{m}$, \cite{PochhammerSymbol}, is used in 
\begin{eqnarray} 
	 \mathcal{K}^m = 2^{m} (\mu + \frac{1}{2} + a)_{m}, 
	 \quad m = 0, 1, 2, \dots,  
\end{eqnarray}
$B^m=\mathcal{K}^m$ when $a=0$ and $A^m=\mathcal{K}^m$ when $a=1$, i.e. $A^m=B^{m+1}/B^1$, and implemented in the \textit{expanded} Eqs. \eqref{eq:1DmuGeneralizedWeights} thereafter. 
Note that Eq. \eqref{eq:1DmuGeneralizedWeights0} is a polynomial in $B \ \theta$ with zeros at $c_1^2, c_2^2,  \dots c_n^2$ (i.e. when $B \ \theta = c_n^2$) and constant term one. For the particular case of $c_1^2=1, c_2^2=q, c_3^2=q^2, \dots, c_z^2=q^{z-1}$ with $q>1$, Eq. \eqref{eq:1DmuGeneralizedWeights0} can be recast in terms of the $q$-Pochhammer symbols $(B \ \theta, 1/q)_z$. A similar analysis can be done for Eq. \eqref{eq:1DmuGeneralizedWeightsk}. 
The $q$-Pochhammer symbol is defined as
\begin{eqnarray} \label{eq:q-PochhammerSymbol}
	(a, q)_z = \prod_{k=0}^{z-1} (1-aq^k), 
\end{eqnarray}
where $(a, q)_0=1$, and reduces to the Pochhammer symbol at the limit $q \rightarrow 1$.  

By definition, the populations are non-negative.  Hence, the weights are non-negative, and thereby the thermal LB model is valid, provided that the $\theta$ is within a range whose extremes (and excluded) values are obtained from the following relations 
\begin{subequations} \label{eq:MaxTemp}
\begin{eqnarray} \label{eq:MaxTempCenterWeights}
\sum_{i=0}^z (-1)^i 2^{z-i} \ (\mu + \frac{1}{2})_{z-i} \ \theta^{z-i}  e_i(c_1^2, c_2^2, \dots, c_z^2) &=& 0,  \\ \label{eq:MaxTempOffCenterWeights}
\sum_{i=0}^{z-1} (-1)^i 2^{z-i} \ (\mu + \frac{1}{2})_{z-i} \ \theta^{z-i}  e_i(c_1^2, c_2^2, \dots, \underbrace{\cancel{c_k^2}}_{\textrm{excluded}}, \dots, c_z^2) &=& 0,  
\end{eqnarray} 
\end{subequations}
which have in turn been obtained from Eqs. \eqref{eq:1DmuGeneralizedWeights0} and \eqref{eq:1DmuGeneralizedWeightsk} respectively, as reformulations by means of $e_i(c_1^2, c_2^2, \dots, c_z^2)$ and $(\mu + \frac{1}{2})_{z-i}$, and equalized to zero. The $e_i(c_1^2, c_2^2, \dots, c_z^2)$ is the $i$th-elementary symmetric polynomial \cite{BatemanVolI-II-III}, \cite{CharalambidesBook}, and $(\mu + \frac{1}{2})_{z-i}$ is the Pochhammer symbol. A recurrent $\theta$ value, obtained from Eq. \eqref{eq:MaxTempOffCenterWeights}, is zero for all $z \geq 1$. The relations \eqref{eq:1DmuGeneralizedWeights} and  \eqref{eq:MaxTemp} have been algebraically computed up to D1Q13 lattice in a general form.  For $n_q >13$ values, particular cases (e.g. with $c_1=1$, $c_2=2$ and so on) can only be tested with today's standard hardware and state of art of symbolic mathematics. Note that for the lattice model D1Q3, then $z=1$ in Eqs. \eqref{eq:MaxTemp} and the theoretical range gives $\theta = ] 0, c_1^2/(2 \mu +1)  [$, which can be reduced to the particular case with $\mu=0$ and $c_1=1$, as it is found in the literature, c.f. \cite{PrasianakisChikatamarlaKarlinAnsumaliBoulouchos2006}. 
Although some weights are never zero or negative for real $\theta$ values with some particular lattice sets, others can become zero or negative under the same conditions. These weights are used to obtain the extremes values of $\theta$ (more about the results on this part is found in section \ref{section:OnTheHigherOrder}, in connection with table \ref{tabular:ValidRangeMuZeroAndGeneralizedTemperatureD1Qnq}).

\subsection{The Equilibrium Distribution Function}\label{section:GeneralizedLatticeBoltzmannEquation}

\begin{figure}[!htbp]
 \begin{center}
\includegraphics[angle=0, width=1.15\textwidth]{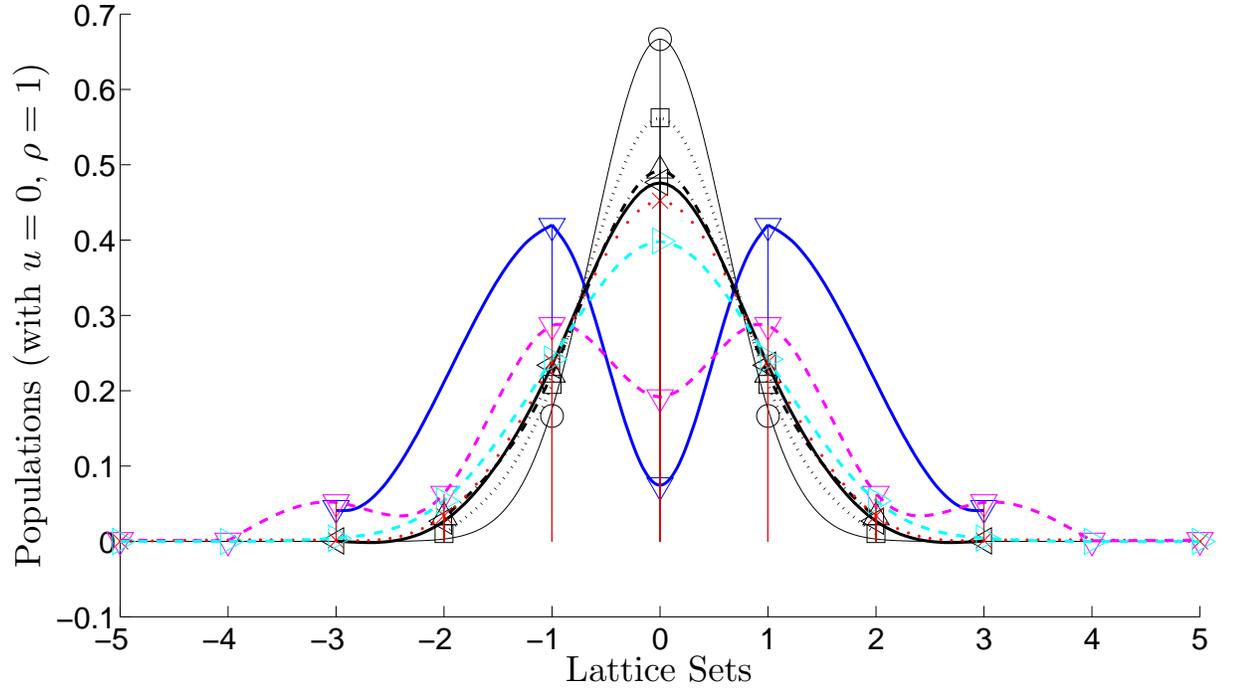}
    \caption{(Color online) Weights values (i.e. populations  $f_i^{\textrm{eq}}$ with $u=0$ and $\rho=1$) and the \textit{likely} shapes of their distributions for the one-dimensional integers lattice set $\boldsymbol{c} = \{  -c_z, \dots -c_1, 0, c_1, \dots, c_z \}$. Symbols are found in table \ref{tab:Weights1DmuGeneralizedHigherOrder}. 
    }
    \label{fig:Weights1DmuGeneralizedHigherOrder}
  \end{center}
\end{figure}

The classical Hermite-based LB construction is derived from a combination between an exponential based weight function and an exponential based equilibrium function \cite{ChenShanPhysicaD2008}. The result leads to the classical EDF $f_i^{\textrm{eq}}=   W_i \sum_{n=0}^{N}  H_n (a)/n ! (b)^n$, where $a=c_{\alpha, i}/ \sqrt{2 \ \theta}$ and $b=u_{\alpha}/ \sqrt{2 \ \theta}$ \cite{ChenShanPhysicaD2008}.  
$M + N \leq Q$, where $Q$ is the degree of precision of the quadrature (c.f. \cite{Shan2010}), $N \geq M$, so that in the low-order LB model $N_{\text{max}}=3$, $M=2$  and thus $Q=5$, is minimum requirement of recovering the Navier-Stokes momentum equation \cite{Shan2010}.   
The generating function for the generalized Hermite polynomial $H_n^{(\mu)}(x)$,  Eq.  \eqref{eq:GeneratingFunctionGeneralizedHermitePolynomial}, suggests   
the introduction of a new equilibrium distribution function, $f_i^{\textrm{eq}}$, i.e. 
\begin{eqnarray} \label{eq1:HermiteMu-EDF}
f_i^{\textrm{eq}} &=& \rho \ W_i \ \sum_{n=0}^{N } \frac{H_n^{(\mu)} (c_{\alpha, i}/\sqrt{2 \ \theta})}{n !} \Bigg(\frac{u_{\alpha}}{\sqrt{2 \ \theta} } \Bigg)^n. 
\end{eqnarray}
For the D1Q3 generalized LB model, with $N=2$ and $N=3$, the result is 
\begin{eqnarray} \nonumber
f_i^{\textrm{eq}} &=& \rho \ W_i \ \Bigg(  1 + \frac{c_{\alpha,i} u_{\alpha}}{\theta (2 \mu +1)} 
- \frac{1}{2} \frac{  u_{\alpha}^2  }{\theta }  + \frac{1}{2} \frac{ c_{\alpha,i}^2 u_{\alpha}^2  }{\theta^2 (2 \mu +1)} \\ \label{eq2:HermiteMu-EDF}
&+& \textcolor{green}{\underline{\textcolor{black}{\frac{1}{2} \frac{ c_{\alpha,i}^3 u_{\alpha}^3}{\theta^3 (2 \mu + 1) (2 \mu + 3)} 
- \frac{1}{2} \frac{ c_{\alpha,i} u_{\alpha}^3  }{\theta^2 (2 \mu +1)}}}}  
\Bigg),
\end{eqnarray}
where the underlined summands correspond to the extra terms due to the $N=3$. Note that the EDF \eqref{eq2:HermiteMu-EDF} is not valid when $\mu=-1/2, -3/2$. The thermal weights  \eqref{eq:1DmuGeneralizedWeights} for the equation model \eqref{eq2:HermiteMu-EDF} are 
\begin{subequations}  \label{eq:ThermalWeightsD1Q3}
\begin{eqnarray} \label{eq:ThermalWeightsD1Q3W0}
		W_0 &=& -\frac{\theta (2 \mu + 1) -c_1^2}{c_1^2}, \\ \label{eq:ThermalWeightsD1Q3W1-2}
		W_{1,2} &=& \frac{1}{2} \frac{\theta (2 \mu + 1) }{c_1^2}.
\end{eqnarray} 
\end{subequations}
Note that the $i$-EDF (c.f. Eqs. \eqref{eq1:HermiteMu-EDF}, \eqref{eq2:HermiteMu-EDF}) equals the $\rho$-scaled $i$-weight ($W_i$) when the lattice flow velocity is zero ($u_{\alpha}=0$). It is easy to see that \underline{with $\mu=0$}, $\theta=c_1^2/3$ and $c_1=1$, Eq.  \eqref{eq2:HermiteMu-EDF} and weights  \eqref{eq:ThermalWeightsD1Q3} are reduced to the classical Hermite-based construction of the low-order lattice Boltzmann formulations, as they are found in the literature, c.f.  \cite{QianHumiereLallemand1992},  \cite{Shan2010}. See also Fig. \ref{fig:Weights1DmuGeneralizedHigherOrder}, where the weights for the D1Q3 model at $\theta=\theta_0$ are represented by the symbol \textcolor{black}{\Large{$\circ$} $-$ } (circle-solid). 

\subsubsection{Model Construction}\label{section:ModelConstruction}

The formulation \eqref{eq2:HermiteMu-EDF}, which contains a free parameter $\mu$, is used in this section. 
The results from the first three classical MB moments, i.e. $\sum_{i=0}^2 f^{\textrm{eq}} c_i^M$ for $M=0, 1, 2$, which corresponds to the density, momentum density and the pressure tensor respectively, are analyzed. 

The density $\sum_{i=0}^{2} f_i^{\textrm{eq}} = \rho$ is fulfilled independently of the value of $\theta$ and $\mu$ for the relation \eqref{eq2:HermiteMu-EDF} and \eqref{eq:ThermalWeightsD1Q3} with both $N=2$ and $N=3$. The momentum density $\sum_{i=0}^{2} f_i^{\textrm{eq}} c_i =j =\rho u$ is also matched under the same conditions for the model with $N=2$, c.f. 	$\textrm{H}_{(2)}^{(0),3}$ and $\textrm{H}_{(2)}^{(\mu),3}$ in table \ref{tab:1DH2H3E}. On the other hand, the momentum density is not fulfilled when $N=3$. 
In the classical Hermite-based construction the issue with $N=3$ is solved with $\theta=\theta_0=1/3$, c.f. $\textrm{H}_{(3)}^{(0),3}$ in table \ref{tab:1DH2H3E}, where $\mu=0$ in \eqref{eq2:HermiteMu-EDF} and \eqref{eq:ThermalWeightsD1Q3}. However, the difference in this work is that both the $\mu$ and $\theta$ can be seen as ``free parameters''. Therefore, the model can be presented with 
\begin{eqnarray} \label{eq:mu-D1Q3}
	\mu = -\frac{1}{2} \frac{3 \theta - c_1^2}{\theta},
\end{eqnarray}
with the condition that $\theta \neq 0$ nor $c_1^2/2$ so that Eqs. \eqref{eq:mu-D1Q3} and \eqref{eq2:HermiteMu-EDF} remain valid respectively, i.e. $\theta = ]0, c_1^2/2[$. Note that with $\theta = \theta_0 = c_1^2/3$ the $\mu=0$ from Eq. \eqref{eq:mu-D1Q3}, which is the known reference ``temperature'' for the low-order (classical) lattice Boltzmann models. 

\begin{table}
	\centering
		{  											
		\begin{tabular}[t]{|ccccccc|}
		\hline
		Eq.  & \vline & $M=1$ & \vline &  $M=2$ & \vline &  $M=3$  \\
 \eqref{eq:MB-convective-moments} & \vline &  & \vline &  & \vline &   \\		
		\hline
		 & \vline & $j_{\alpha}=$ & \vline & $P_{\alpha \alpha}=$ & \vline & $Q_{\alpha \alpha \alpha}=$  \\
		 & \vline & \textrm{{\normalsize $\sum_i$}} $f_i  c_{\alpha,i}$ & \vline & 
		 \textrm{{\normalsize $\sum_i$}} $f_i  c_{\alpha,i}^2$ 
		 & \vline &  \textrm{{\normalsize $\sum_i$}} $f_i  c_{\alpha,i}^3$  \\
		\hline
		\hline
				$\textrm{H}_{(2)}^{(0),3}$  & \vline & $\rho u_{\alpha}$ &  \vline & $\textcolor{blue}{\underline{\textcolor{black}{\rho (\theta_0 + u_{\alpha}^2 )}}}$ &  \vline & $\textcolor{blue}{\uwave{\textcolor{black}{  \rho  u_{\alpha}  }}} + 0 \cdot \textcolor{red}{\underline{\textcolor{black}{  \textcolor{red}{\underline{\textcolor{black}{   \rho u_{\alpha}^3   }}}   }}}$    \\
		
		\hline
				$\textrm{H}_{(3)}^{(0),3}$  & \vline & $\textcolor{blue}{\underline{\textcolor{black}{\rho u_{\alpha}}}}$ & \vline & $\textcolor{blue}{\underline{\textcolor{black}{\rho (\theta_0 + u_{\alpha}^2 )}}}$ &  \vline & $\textcolor{blue}{\uwave{\underline{\textcolor{black}{  \rho  u_{\alpha}  }}}} + 0 \cdot \textcolor{red}{\underline{\textcolor{black}{  \textcolor{red}{\underline{\textcolor{black}{   \rho u_{\alpha}^3   }}}   }}}$    \\
\hline
					$\textrm{H}_{(2)}^{(\mu),3}$  & \vline & $\rho u_{\alpha}$ &  \vline & \textcolor{cyan}{\fbox{ \textcolor{black}{ $\rho (c_1^2 - 2 \theta) + \rho u_{\alpha}^2$  } }}  &  \vline & $\textcolor{blue}{\uwave{\textcolor{black}{  \rho  u_{\alpha}  }}} + 0 \cdot \textcolor{red}{\underline{\textcolor{black}{  \textcolor{red}{\underline{\textcolor{black}{   \rho u_{\alpha}^3   }}}   }}}$    \\

		\hline				
		$\textrm{H}_{(3)}^{(\mu),3}$  & \vline & \textcolor{cyan}{\fbox{ \textcolor{black}{ $\rho u_{\alpha}$  } }} & \vline & \textcolor{cyan}{\fbox{ \textcolor{black}{ $\rho (c_1^2 - 2 \theta) + \rho u_{\alpha}^2$  } }}  &  \vline & \textcolor{cyan}{\fbox{ \textcolor{black}{ \textcolor{blue}{\uwave{\textcolor{black}{  $\rho  u_{\alpha}$  }}} } }} $+ 0 \cdot \textcolor{red}{\underline{\textcolor{black}{  \textcolor{red}{\underline{\textcolor{black}{   \rho u_{\alpha}^3   }}}   }}}$    \\
	
	\hline
	
			$\textrm{E}_{(1)}^3$  & \vline & $\textcolor{blue}{\underline{\textcolor{black}{\rho u_{\alpha}}}}$ & \vline &  $\textcolor{blue}{\underline{\textcolor{black}{\rho ( \theta_0 + u_{\alpha}^2 )}}}$ & \vline &  $\textcolor{blue}{\uwave{\underline{\textcolor{black}{  \rho  u_{\alpha}  }}}} + 0 \cdot \textcolor{red}{\underline{\textcolor{black}{  \textcolor{red}{\underline{\textcolor{black}{   \rho u_{\alpha}^3   }}}   }}}$    \\
	 & \vline & & \vline &   \textcolor{red}{\fbox{\fbox{ \textcolor{blue}{\underline{\textcolor{black}{ $+ \mathcal{O}(u^4) $ }}}   }}}  & \vline &  \\
											
		\hline	
		$\textrm{E}_{(2)}^3$ & \vline & $j_{\alpha}$ & \vline & $\rho \mathbb{P}_{\alpha \alpha}$ & \vline & $\textcolor{blue}{\uwave{\textcolor{black}{  j_{\alpha}  }}} + 0 \cdot \textcolor{red}{\underline{\textcolor{black}{  \textcolor{red}{\underline{\textcolor{black}{   \rho u_{\alpha}^3   }}}   }}}$    \\ 
				\hline	  		
		\end{tabular}
		}
			\caption{(Color online) Comparison among the ``entropic'',  classical and the $\mu$-generalized Hermite-based one-dimensional lattice LB models, where $f_i$ is $f_i^{\textrm{eq}}$. The $\alpha$ is the coordinate axis.  $\textrm{H}_{(N)}^{(\mu),n_q}$: results from the (classical or $\mu$-generalized) Hermite-based construction Eq. \eqref{eq1:HermiteMu-EDF} with $N=2$ or $N=3$ using a number of discrete lattice velocity vectors $n_q$ and $\mu =0 $ or Eq.  \eqref{eq:mu-D1Q3}. 
			$\textrm{E}_{(M_{\textrm{max}})}^{n_q}$: results from the ELB construction (c.f. appendix). 
			The matching terms to the classical Maxwell-Boltzmann (MB) moments are not underlined or under-wave. The single underlined terms are conditioned to $\theta_0 = c_1^2/3$, while the under-wave to $c_1=1$. The missing MB terms are double underlined. The terms within a box are conditioned to Eq. \eqref{eq:mu-D1Q3}. The double boxed summands are spurious terms. 
			The term $Q_{\alpha \alpha \alpha}$ matches its corresponding MB moment at low Mach number provided $\theta = \theta_0 = c_1^2/3$, $ c_1=1$ and $u_{\alpha}^3 \approx 0$.
The mass conservation ($\rho$, not shown in the table) is achieved  for all the models.}
	\label{tab:1DH2H3E}
\end{table}

The resulting terms $\rho u_{\alpha}$  for both with $M=1$ and $M=3$, corresponding to the construction $\textrm{H}_{(3)}^{(\mu),3}$, c.f. table \ref{tab:1DH2H3E}, are obtained under similar thermal $\theta$ conditions as in $\textrm{H}_{(2)}^{(0),3}$,  when the generalized Hermite-based LB construction is introduced. On the other hand, the relation \eqref{eq:mu-D1Q3} has no effect on the same terms for the $\textrm{H}_{(2)}^{(\mu),3}$ construction. This is an  (algebraic) improvement over the classical Hermite-based LB construction. 
From  the results in table \ref{tab:1DH2H3E} for $M=2$, $\textrm{H}_{(2)}^{(\mu),3}$ and $\textrm{H}_{(3)}^{(\mu),3}$, and the inviscid momentum flux density \cite{LandauLifshitzPKBook} (c.f. Eq. (5.11) therein), the pressure $p=\rho (c_1^2 - 2 \theta )$  is identified. The lattice ``speed of sound'' 
 yields 
\begin{eqnarray} \nonumber
	c_{\textrm{sound}} &=& \sqrt{ \frac{\partial p}{\partial \rho} } \\ \label{eq:SpeedOfSound-muLowOrderLB}
										 &=&	\sqrt{ c_1^2 - 2 \theta }.
\end{eqnarray}
The value of so called reference ``temperature'' $\theta=\theta_0=c_1^2/3$ is required in the low-order classical Hermite-based LB constructions $\textrm{H}_{(2)}^{(0),3}$ and $\textrm{H}_{(3)}^{(0),3}$ \cite{QianHumiereLallemand1992}. 
This eliminates spurious velocity terms in their pressure tensors, which are thereby matched to the classical MB moment $M=2$  isothermally, c.f. table \ref{tab:1DH2H3E}. On the other hand, the value of $\theta$ is found in the $\mu$-generalized Hermite-based LB constructions  $\textrm{H}_{(2)}^{(\mu),3}$ and $\textrm{H}_{(3)}^{(\mu),3}$ and no spurious velocity terms are seen in table  \ref{tab:1DH2H3E}. Note that with $\theta = \theta_0 = c_1^2/3$, the Eq. \eqref{eq:SpeedOfSound-muLowOrderLB} is reduced to $c_{\textrm{sound}}=\sqrt{c_1^2/3}$.

It is convenient to recall at this point that the \textit{physical} speed of sound $c_{\textrm{sound}}=\sqrt{ \gamma \theta}$ and thus, a comparison with the classical lattice  $c_{\textrm{sound}}=\sqrt{ \theta}$ implies that $\theta=\gamma \theta$, i.e. $\theta (1-\gamma)=0$, where $\theta \neq 0$ so that Eq. \eqref{eq2:HermiteMu-EDF} is valid, regardless $\mu$. Hence, $\gamma = 1 + 2/D_{\textrm{m}}=1$, i.e. the degree of freedom of molecules $D_{\textrm{m}} = \infty$, which is unphysical. 
This leads to the Newton's speed of sound $c_{\textrm{sound}}=\sqrt{ \theta}$, which uses the ideal gas equation of state 
$p=\rho \theta$ found in the Euler equation and $\theta = $ constant, i.e. isothermal assumption. $p=\rho \theta$ is found in  the pressure tensor, obtained from the MB moment $M=2$ in \eqref{eq:MB-convective-moments} (more about this below). 
On the other hand, based on \eqref{eq:SpeedOfSound-muLowOrderLB} the result is $c_1^2 - 2 \theta= \gamma \theta$, which yields
\begin{eqnarray} \label{eq:ReferenceTemperatureMuHLB}
	\theta=\frac{c_1^2}{\gamma + 2}=\frac{ D_{\textrm{m}} c_1^2}{3 D_{\textrm{m}} +2}.
\end{eqnarray}
Because there is no any restrictions on the $\theta$ value for the moments $M=0$ and 1 terms, c.f. $\textrm{H}_{(2)}^{(\mu),3}$ and $\textrm{H}_{(3)}^{(\mu),3}$ in table \ref{tab:1DH2H3E}, then 
the $\theta$ values in \eqref{eq:ReferenceTemperatureMuHLB}, for monoatomic molecules ($D_{\textrm{m}}=3, \gamma=5/3$) and diatomic molecules ($D_{\textrm{m}}=5, \gamma=7/5$), can be $\theta = 3 c_1^2/11$ and $\theta = 5 c_1^2/17$ respectively.  
However, $\theta = \theta_0 = c_1^2/3$ and $c_1=1$ are required in the term $Q_{\alpha \alpha \alpha}$ , c.f. table \ref{tab:1DH2H3E}, so that the classical MB moment $M=3$ is matched when $u_{\alpha}^3 \approx 0$. 
This limitation on the $Q_{\alpha \alpha \alpha}$ term is imposed on \textit{all} low-order LB models found in table \ref{tab:1DH2H3E}, where the ``entropic'' (c.f. appendix),  classical and the $\mu$-generalized Hermite-based one-dimensional LB models are included. 
The strategy of having a deviation around $\theta_0$, 
as in \cite{PrasianakisChikatamarlaKarlinAnsumaliBoulouchos2006}, at the expense of the accuracy of the MB moment $M=3$, has found no 
applications among practitioners dealing with weakly compressible flows to the best knowledge of the author. 

The results show so far that a fixed value of $\theta=\theta_0$ is required to achieve the \textit{best possible} accuracy to reconstruct the incompressible  Navier-Stokes equation from the low-order LB models, i.e. with $z=1$, in the low Mach-number limit when the models are free of correcting counter terms and regardless the $\mu$ value. However, this can be changed when $z>1$, depending on the LB construction approach. For example, when $z \geq 3$ in Eqs. \eqref{eq:1DmuGeneralizedWeights} and $N \geq 4$ in Eq. \eqref{eq1:HermiteMu-EDF} with $\mu \neq 1/2 - n$, $n=1/2, 1, 2, 3, \dots$ the $\sum_{i=0}^{n_q-1} f_i^{\textrm{eq}} c_i^M $ for $M=0, 1, 2$ and 3 gives $\rho$, $j=\rho u$,  $P=\rho \theta (1 +  2 \mu) + \rho u^2$ and $Q=\rho \theta u (3 +  2 \mu) + \rho u^3$  respectively. That is, a \textit{new} 
moment system (to be denoted as $\mathcal{M}$) is obtained with the use of the proposed  $\mu$-generalized Hermite-based LB construction approach, from which the classical MB moment system is a particular case with $\mu=0$. The area of application of this $\mathcal{M}$ moment system will be determined by what is wanted to be achieved.  Anyhow, it is already noticed here that based on an approach to obtain the macroscopic relations (e.g. method of moments) on the LBGK equation, the solution of  
\begin{eqnarray} 
\partial_t j  + \partial_x P + \partial_x \Big(- \Big(\tau-\frac{1}{2}\Big)\Big( \partial_t P + \partial_x Q  \Big) \Big) = 0, 
\end{eqnarray}
with constant $\theta$ yields the one-dimensional Navier-Stokes equation 
\begin{eqnarray} \label{eq:IsothermalNavier-Stokes}
\partial_t j  + \partial_x (p + \rho u^2) - \partial_x ( 2 \rho \nu \partial_x u)  = 0, 
\end{eqnarray}
where $p=\rho c_{\textrm{sound}}^2$ and $\nu = (\tau-1/2) c_{\textrm{sound}}^2$. When the MB moments are used,  $c_{\textrm{sound}}=\sqrt{ \theta}$ is obtained. On the other hand, $ c_{\textrm{sound}} = \sqrt{ (1 + 2 \mu) \theta }$ is obtained when the aforementioned new $\mathcal{M}$ moment system is implemented. 
Although it is noted that the value of $\mu=1/D_{\textrm{m}} $ can be extracted from $\gamma \theta = (1 + 2 \mu)  \theta $,  
the existence of a free $\mu$ parameter can be useful when dealing with ($\textit{on}$-Cartesian) lattice sizes. 
More about these $\mathcal{M}$ moments in section \ref{section:OnTheHigherOrder}.


%

\newpage

\begin{table} [!htbp]
	\centering
		{  											
		\begin{tabular}[t]{|ccccccc|}
		\hline
		 Symbols in Fig. \ref{fig:Weights1DmuGeneralizedHigherOrder}& \vline & Model  & \vline & $\theta$ & \vline &  $\boldsymbol{c}$  \\ 
		\hline
		 \textcolor{black}{\Large{$\circ$} $-$ }  & \vline & D1Q3 (with Eq. \eqref{eq:mu-D1Q3}) & \vline &  $\theta =c_1^2/3$  & \vline &  $c_1=1$ \\
		  (circle-solid) & \vline & (z=1)  & \vline &  & \vline &   \\
		\hline
		 \textcolor{black}{$\square$ $\cdot$ $\cdot$}    & \vline & D1Q5 with $\mu=0$ & \vline &  $\theta =0.5$  & \vline &  $c_1=1$ \\
		  (squared-dotted) & \vline & (z=2)  & \vline &  & \vline &  $c_2=2$ \\
		  \hline
		 \textcolor{black}{\Large{$\vartriangle$} -. }    & \vline & D1Q5 with $\mu=0$ & \vline &  $\theta =0.7$  & \vline &  $c_1=1$ \\
		  (triangle up-dashdot) & \vline & (z=2)  & \vline &  & \vline &  $c_2=2$ \\
		\hline	
		 \textcolor{blue}{\Large{$\triangledown$} $-$}   & \vline & D1Q5 with $\mu=0$ & \vline &  Eq.  \eqref{eq:ReferenceTemperatureD1Q5}  & \vline &  $c_1=1$ \\
		  (triangle down-solid) & \vline & (z=2)  & \vline &  c.f. Eq. \ref{eq:RefertenceTemperatureD1Q5WithC2-3} & \vline &  $c_2=3$ \\
		\hline		
		\textcolor{black}{\Large{$\triangleleft$} $-$}  & \vline & D1Q7 with $\mu=0$ & \vline &  Eq.  \eqref{eq:ReferenceTemperatureD1Q7} & \vline &  $c_1=1$ \\
		  (triangle left-solid) & \vline & (z=3)  & \vline & c.f. Eq. \ref{eq:RefertenceTemperatureD1Q7WithC3-3} & \vline &  $c_{2,3}={2,3}$  \\
		\hline	
				\textcolor{red}{$\times$ $\cdot$ $\cdot$}  & \vline & D1Q9 with $\mu=0$ & \vline &  Eq.  \eqref{eq:RefertenceTemperatureD1Q9WithC5} & \vline &  $c_{1,2}={1,2}$ \\
		  (cross-dotted) & \vline & (z=4)  & \vline &  & \vline &  $c_{3,4}={3,5}$  \\
			\hline
			 \textcolor{cyan}{\Large{$\triangleright$} - -}  & \vline & D1Q11 with $\mu=0$ & \vline &  $\theta=1.0$ & \vline &  $c_{1,2}={1,2}$ \\
		  (triangle right-dashed) & \vline & (z=5)  & \vline &  & \vline &  $c_{3,4,5}={3,4,5}$  \\
			\hline	
		 \textcolor{magenta}{\Large{$\triangledown$} - -}  & \vline & D1Q11 with $\mu=0$ & \vline &  Eq. \eqref{eq:RefertenceTemperatureD1Q11WithC5-5} & \vline &  $c_{1,2}={1,2}$ \\
		  (triangle down-dashed) & \vline & (z=5)  & \vline &  & \vline &   $c_{3,4,5}={3,4,5}$   \\
			\hline			
		\end{tabular}
		}
			\caption{(Color online) Symbols corresponding to Fig. \ref{fig:Weights1DmuGeneralizedHigherOrder}.}
	\label{tab:Weights1DmuGeneralizedHigherOrder}
\end{table}

\newpage

\section{On The High-Order LB Model}\label{section:OnTheHigherOrder}

The second issue to treat in this work is whether or not there exist Hermite-based high-order lattice Boltzmann D1Q$n_q$ models, $n_q \geq 5$, so that they are able to fulfill the following three characteristics within a \textit{single} construction: 
capable to exactly match the first hydrodynamic $z$-moments with free $\theta$ values   
1) \textit{on}-Cartesian, 2) with \textit{thermal} weights (based on the final results that are used in the EDF), 3) 
whilst the hydrodynamic $(z+1)$-moments are exactly matched with the \textit{shortest} \textit{on}-Cartesian lattice sets with some fixed $\theta$ values. 
Most of the existing high-order LB models are based the classical MB moment system. Because comparisons are made in this section, the final results are presented here \textit{first} for the particular case when $\mu=0$. The case when $\mu \neq 0$  is studied subsequently.  
Although Eq. \eqref{eq1:HermiteMu-EDF} becomes the same relation as the one in \cite{ChenShanPhysicaD2008} when $\mu$ is set to zero, the proposed formulation of the weights \eqref{eq:1DmuGeneralizedWeights}, used directly in the final EDF, are still thermal, unlike those athermal used/obtained in \cite{ShanYuanChen2006} (tables 1,2,3 therein), \cite{ShanChen2007} (table 1 therein), \cite{MengZhangJCP2011} (table 1 therein), just to mention few examples. Furthermore, the lattice $c_k$ values in the thermal weights \eqref{eq:1DmuGeneralizedWeights} can be integers.

It can be shown that the results of combining the thermal weights \eqref{eq:1DmuGeneralizedWeights}  and the relation \eqref{eq1:HermiteMu-EDF} 
 with $\mu=0$ leads to that the MB $M$-moments can be thermally matched, i.e. with $\sum f_i^{\textrm{eq}} c_{\alpha,i}^M$, in the D1Q$n_q$ models up to $M=z$. $M=0, 1, 2 \dots$, and $n_q=3, 5, 7 \dots$. The MB $(z+1)$-moment is completely matched solely at a certain fixed reference $\theta=\theta_0$ value. Alternatively, the MB $(z+1)$-moment can be thermally matched up to the velocity term $u^{z-1}$. The rest of the higher order $\mathcal{Z}$-moments, where $\mathcal{Z}>(z+1)$, are not completely guaranteed. 
The implemented Hermite $N = 3, 4, 5, 6, \dots$ order in \eqref{eq1:HermiteMu-EDF} are for the one-dimensional $n_q = 5, 7, 9, 11, \dots$ respectively. The aforementioned fixed values of $\theta_0$ can be obtained, e.g. for the D1Q3, D1Q5, D1Q7 and D1Q9 models, from the relation 
\begin{eqnarray} \label{eq:ReferenceTemperatureD1Qnq} 
&&\sum_{k=0}^z (-1)^{k}(n_q-2 \cdot k)!! \ \theta^{z-k} e_k(c_1^2, c_2^2, \dots, c_z^2) = 0, \\ \nonumber
\end{eqnarray} 
while for the D1Q11 model, the term $- 540$ is added into the $\theta^5$ part of the polynomial generated by \eqref{eq:ReferenceTemperatureD1Qnq}, from which the roots are obtained.  
Hence, bulky expressions are avoided in the present work. $e_k(c_1^2, c_2^2, \dots, c_z^2)$ is a $k$th-elementary symmetric polynomial. $k$ are non-negative integer numbers. 

Some examples are outlined to corroborate the aforementioned statements. The results for D1Q5 LB model are 
\begin{subequations} \label{eq:AccuracyD1Q51D}
\begin{eqnarray} \label{eq:AccuracyD1Q51D-0}
\sum_{i=0}^{n_q-1} f_i^{\textrm{eq}} c_i^0 &=& \rho,  \\ \label{eq:AccuracyD1Q51D-1}
\sum_{i=0}^{n_q-1} f_i^{\textrm{eq}} c_i^1 &=& \rho u,  \\  \label{eq:AccuracyD1Q51D-2}
\sum_{i=0}^{n_q-1} f_i^{\textrm{eq}} c_i^2 &=& \rho (\theta + u^2),  \\ \label{eq:AccuracyD1Q51D-3}
Q^{\textrm{eq}} = \sum_{i=0}^{n_q-1} f_i^{\textrm{eq}} c_i^3 &=& \rho  (Q_1 \theta u +  Q_3 u^3),  \\ \label{eq:AccuracyD1Q51D-4}
R^{\textrm{eq}} = \sum_{i=0}^{n_q-1} f_i^{\textrm{eq}} c_i^4 &=& \rho (R_0 \theta^2 + R_2 \theta u^2 + R_4 u^4),   \\ \label{eq:AccuracyD1Q51D-5}
S^{\textrm{eq}} = \sum_{i=0}^{n_q-1} f_i^{\textrm{eq}} c_i^5 &=& \rho ( S_1 \theta^2 u +  S_3 \theta u^3 + S_5 u^5), 
\end{eqnarray}
\end{subequations}
where Eqs. \eqref{eq:AccuracyD1Q51D-0}-\eqref{eq:AccuracyD1Q51D-4} represent the density, momentum density, pressure tensor, energy flux and the rate of change of the energy flux respectively. The $Q_i$, $R_i$ and $S_i$ are the MB coefficients. $Q_1=3$ and $R_0=3$ in Eqs. \eqref{eq:AccuracyD1Q51D-3} and \eqref{eq:AccuracyD1Q51D-4} respectively. Their values for the lattice set D1Q5 model are:
\begin{eqnarray} \label{eq:MBcoefficientForD1Q5-Q3}
Q_3 &=& \frac{e_2(c_1^2,c_2^2,- 3 \theta) }{-6 \theta^2} - \frac{3}{2} 
\\ \label{eq:MBcoefficientForD1Q5-R2}
R_2 &=& \frac{e_2(c_1^2,c_2^2,- 3 \theta) }{- 2 \theta^2} - \frac{3}{2}, 
\\  \label{eq:MBcoefficientForD1Q5-S1}
S_1 &=& \frac{e_2(c_1^2,c_2^2,- 3 \theta)}{-\theta^2}, 
\\ \label{eq:MBcoefficientForD1Q5-S3}
S_3 &=& - \frac{c_1^2 c_2^4 + 9 c_1^2 \theta^2 -3 c_1^4 \theta + c_1^4 c_2^2 - 3 c_2^4 \theta - 6 \theta c_1^2 c_2^2 + 9 c_2^2 \theta^2}{6 \theta^3}, \\ \label{eq:MBcoefficientForD1Q5-R4}  
R_4 &=& 0, \\ \label{eq:MBcoefficientForD1Q5-S5}
S_5 &=& 0.
\end{eqnarray} 
Complete Galilean invariant is achieved when $Q_3=1$ (c.f. the relation  \eqref{eq:AccuracyD1Q51D-3}). Therefore, from Eq. \eqref{eq:MBcoefficientForD1Q5-Q3} yields
\begin{eqnarray} \label{eq:ReferenceTemperatureD1Q5}
\theta = \theta_0 = \frac{c_1^2+c_2^2}{10} + \frac{\sqrt{(3 c_2^2 - 3 c_1^2)^2-24 c_1^2 c_2^2}}{30},   
\end{eqnarray} 
which is a particular case of Eq. \eqref{eq:ReferenceTemperatureD1Qnq} with $z=2$.
Hence, $c_1 =1$ and $c_2 \neq 2$, otherwise the reference ``temperature'' $\theta_0$ \eqref{eq:ReferenceTemperatureD1Q5} is complex-valued with $c_2 = 2$. With $c_1 =1$ and $c_2 =3$ in Eq. \eqref{eq:ReferenceTemperatureD1Q5} yields  $\theta_0=1+\sqrt{10}/5$. Although the MB $(z+1)$-moment is matched isothermally with $z=2$  for the lattice set D1Q5 model, the energy flux is partially fulfilled thermally for low Mach number provided that $u^3 \approx 0$ can be assumed. The rest of the MB coefficients \eqref{eq:MBcoefficientForD1Q5-R2}-\eqref{eq:MBcoefficientForD1Q5-S3} can be obtained with $\theta$ from Eq.  \eqref{eq:ReferenceTemperatureD1Q5}, $c_1 =1$ and $c_2 =3$, leading to $R_2=6=R_2^{\textrm{MB}}$,  $S_1=1=S_1^{\textrm{MB}}$ and $S_3=250 (7 + 2 \sqrt{2 \cdot 5})/(5+\sqrt{2 \cdot 5})^3 \approx 6.1257 \neq  S_3^{\textrm{MB}}=10$. Note that $R_4 = 0 \neq R_4^{\textrm{MB}}  = 1$ and $S_5 =0 \neq S_5^{\textrm{MB}}  = 1$ regardless the values of $c_1$ and $c_2$, i.e. they are unconditioned no matching term to the MB coefficients.

For the D1Q7 model,  the results for the $\sum_{i=0}^{n_q-1} f_i^{\textrm{eq}} c^M$ moments $M = 0, 1, 2$ are the same as the relations  \eqref{eq:AccuracyD1Q51D-0}-\eqref{eq:AccuracyD1Q51D-2}, while the rest are
\begin{subequations} \label{eq:AccuracyD1Q71D}
\begin{eqnarray} \label{eq:AccuracyD1Q71D-0}
Q^{\textrm{eq}} = \sum_{i=0}^{n_q-1} f_i^{\textrm{eq}} c_i^3 &=& \rho (3 \theta u + u^3 ), \\ \label{eq:AccuracyD1Q71D-4}
R^{\textrm{eq}} = \sum_{i=0}^{n_q-1} f_i^{\textrm{eq}} c_i^4 &=& \rho (3 \theta^2 + 6 \theta u^2 + R_4 u^4), \\ \label{eq:AccuracyD1Q71D-5}
S^{\textrm{eq}} = \sum_{i=0}^{n_q-1} f_i^{\textrm{eq}} c_i^5 &=& \rho ( 15 \theta^2 u +  S_3 \theta u^3 + S_5 u^5), \\ \label{eq:AccuracyD1Q71D-6}
V^{\textrm{eq}} = \sum_{i=0}^{n_q-1} f_i^{\textrm{eq}} c_i^6 &=& \rho ( V_0 \theta^3  +  V_2 \theta^2 u^2 + V_4 \theta u^4 + V_6 u^6), 
\end{eqnarray}
\end{subequations}
i.e. it is complete Galilean invariant \textit{thermally}. $V_0=15$ in Eq. \eqref{eq:AccuracyD1Q71D-6}. The MB coefficients for the lattice set D1Q7 model are:
\begin{eqnarray} \label{eq:MBcoefficientForD1Q7-R4}
R_4 &=& \frac{e_3(c_1^2,c_2^2,c_3^2,- 3 \theta)+ 15 \theta^2 e_1(c_1^2,c_2^2,c_3^2)- 81 \theta^3}{24 \theta^3}, \\ \label{eq:MBcoefficientForD1Q7-S3}
S_3 &=& \frac{e_3(c_1^2,c_2^2,c_3^2,- 3 \theta)+ 15 \theta^2 e_1(c_1^2,c_2^2,c_3^2)- 45 \theta^3}{6 \theta^3},  \\ \label{eq:MBcoefficientForD1Q7-S5}
S_5 &=& 0, \\ \label{eq:MBcoefficientForD1Q7-V2}
V_2 &=& \frac{e_3(c_1^2,c_2^2,c_3^2,- 3 \theta)+ 15 \theta^2 e_1(c_1^2,c_2^2,c_3^2)- 15 \theta^3}{2 \theta^3}, \\ \label{eq:MBcoefficientForD1Q7-V4}
V_4 &=&\frac{1}{24 \theta^4} \Big( -90 c_{2}^2 \theta^3-12 c_{1}^2 \theta c_{2}^2 c_{3}^2-3 c_{3}^2 c_{1}^4 \theta-3 c_{3}^4 c_{1}^2 \theta-3 c_{3}^4 c_{2}^2 \theta \\ \nonumber
&&+c_{1}^4 c_{3}^2 c_{2}^2+c_{2}^2 c_{1}^2 c_{3}^4-3 c_{1}^4 \theta c_{2}^2+33 c_{2}^2 \theta^2 c_{1}^2+33 c_{2}^2 \theta^2 c_{3}^2+33 c_{1}^2 \theta^2 c_{3}^2 \\ \nonumber
&&+15 \theta^2 c_{2}^4-3 c_{3}^2 \theta c_{2}^4-3 c_{1}^2 \theta c_{2}^4+c_{1}^2 c_{3}^2 c_{2}^4+45 \theta^4+15 c_{1}^4 \theta^2 \\ \nonumber
&& -90 \theta^3 c_{3}^2-90 \theta^3 c_{1}^2+15 c_{3}^4 \theta^2 \Big), \\ \label{eq:MBcoefficientForD1Q7-V6}
V_6 &=& 0.
\end{eqnarray} 
The rate of change of the energy flux is completely fulfilled when $R_4 = 1 $ in Eq. \eqref{eq:AccuracyD1Q71D-4}. Hence, a reference value of $\theta=\theta_0$ is derived from Eq. \eqref{eq:MBcoefficientForD1Q7-R4}, which is a particular case of the relation \eqref{eq:ReferenceTemperatureD1Qnq} with $z=3$. With $c_1 =1$, $c_2 =2$, $c_3 =3$ and $z=3$, Eq. \eqref{eq:ReferenceTemperatureD1Qnq} becomes  
\begin{eqnarray} \label{eq:ReferenceTemperatureD1Q7}
\theta_0 &=& \frac{1}{150} \frac{(1225+735 \sqrt{30})^{2/3} -245+70(1225+735 \sqrt{30})^{1/3}}{(1225+735 \sqrt{30})^{1/3}}. 
\end{eqnarray} 
Complex values of $\theta_0$ appears in the D1Q7 model too, for instance, with $c_1 =1$, $c_2 =3$ and $c_3 \geq 5$, which can be avoided using combinations such as $c_1 =1$, $c_2 =2$ and $c_3 = 3$ or 4. 

The values of $\theta_0$ for some lattice set cases are found in table \ref{tabular:ReferenceTemperatureD1Qnq}, which are presented as large numbers (around machine precision) for the sake of compassion to \cite{ChikatamarlaKarlin2009}. The accuracy of the MB coefficients conditioned to $\theta = \theta_0$ is proportional to the accuracy of the $\theta_0$ value. The MB coefficients for the lattice set D1Q$n_q$ models with $n_q=5, 7, 9$ and 11 are summed up in table  \ref{tab:MBCoefficientsD1Qnq}. From the D1Q5 and D1Q7 results, the $u^{z+1}$ velocity terms of the MB coefficients belonging to the MB $(z+3)$ moments, i.e. $S_3$ and $V_4$ respectively, are closest to their MB values when integers lattice $c_k$ are used  if the $\theta_0$ is obtained using a lattice velocity set $\boldsymbol{c}$, which is as short as possible (c.f. $\theta_0$ values in table \ref{tabular:ReferenceTemperatureD1Qnq}), . For instance, $S_3 \approx 6.1257$, $S_3 \approx 5.5732$ and $S_3 \approx 5.3532$ are obtained for the D1Q5 model with $\theta_0$ computed with the lattice sets $\{0, \pm 1, \pm 3 \}$ (c.f. \eqref{eq:RefertenceTemperatureD1Q5WithC2-3} in table \ref{tabular:ReferenceTemperatureD1Qnq}), $\{0, \pm 1, \pm 4 \}$ and $\{ 0, \pm 1, \pm 5 \}$ respectively. The last two lattice sets give negative weights. 
$V_4 \approx 13.7497$, $V_4 \approx 10.4718$, $V_4 \approx 8.8344$ and $V_4 \approx -124.9263$ are obtained for the D1Q7 model with $\theta_0$ computed with the lattice sets $\{0, \pm 1, \pm 2, \pm 3 \}$ (c.f. \eqref{eq:RefertenceTemperatureD1Q7WithC3-3} in table \ref{tabular:ReferenceTemperatureD1Qnq}), $\{0, \pm 1, \pm 2, \pm (4-5) \}$ and $\{0, \pm 1, \pm 3, \pm 4 \}$ respectively. The last two lattice sets give negative weights.  
The rest of the MB coefficients remain the same as they are presented in table \ref{tab:MBCoefficientsD1Qnq} when the $\theta_0$ values in table \ref{tabular:ReferenceTemperatureD1Qnq} are implemented for these two D1Q5 and D1Q7 models (more about this below).    

\begin{table}[!htbp]
	\centering
	{ \footnotesize
\begin{tabular}{|p{0.95\linewidth}|} 
\hline 
\begin{subequations} \label{eq:RefertenceTemperatureD1Q5With012}
			\begin{eqnarray} \label{eq:RefertenceTemperatureD1Q5WithC2-2}
			\theta_0( \textrm{ with } \boldsymbol{c}=\{0, \pm 1, \pm 2	 \} ) &=& \textrm{complex valued}, \\ \label{eq:RefertenceTemperatureD1Q5WithC2-3}
			\theta_0( \textrm{ with } \boldsymbol{c}=\{0, \pm 1, \pm 3	 \} ) &=& 
			 1+\frac{\sqrt{10}}{5}.
			\end{eqnarray}
\end{subequations} \\
\hline 	 
			\begin{eqnarray} \label{eq:RefertenceTemperatureD1Q7WithC3-3}
			\theta_0( \textrm{ with } \boldsymbol{c}=\{0, \pm 1, \pm 2	\pm 3	 \} ) &=& 
			0.697 \ 953 \ 322 \ 019 \ 683 \ 088 \ 24.
			\end{eqnarray}
\\
\hline 
\begin{subequations} \label{eq:RefertenceTemperatureD1Q9WithC4-5-8}
\begin{eqnarray} \label{eq:RefertenceTemperatureD1Q9WithC4}
	\theta_0 (\textrm{ with } \boldsymbol{c}=\{0, \pm 1, \pm 2, \pm 3, \pm 4	 \} ) &=& \textrm{complex valued}, \\ \label{eq:RefertenceTemperatureD1Q9WithC5}
	\theta_0 (\textrm{ with } \boldsymbol{c}=\{0, \pm 1, \pm 2, \pm 3, \pm 5	 \} ) &=& 
	0.756 \ 080 \ 852 \ 594 \ 268 \ 582 \ 31.   
 \end{eqnarray}
\end{subequations} 
\\
\hline  
\begin{eqnarray} \label{eq:RefertenceTemperatureD1Q11WithC5-5}
	\theta_0 (\textrm{ with } \boldsymbol{c}=\{0, \pm 1, \pm 2, \pm 3, \pm 4, \pm 5	 \} ) &=& 
	2.123 \ 517 \ 542 \ 924 \ 955 \ 553 \ 8.   
\end{eqnarray}
\\
\hline 
\end{tabular}
	}
\caption{Reference values of $\theta=\theta_0$ for some one-dimensional  D1Q$n_q$ models and 
shortest \textit{on}-Cartesian lattice velocity sets $\boldsymbol{c} = \{0, \pm c_1, \dots, \pm c_z \}$, so the MB  $(z+1)$-moment is completely fulfilled, c.f. Eq.  \eqref{eq:ReferenceTemperatureD1Qnq}.  Eq. \eqref{eq:RefertenceTemperatureD1Q5WithC2-3}: D1Q5; Eq. \eqref{eq:RefertenceTemperatureD1Q7WithC3-3}: D1Q7; Eq. \eqref{eq:RefertenceTemperatureD1Q9WithC5}: D1Q9; Eq. \eqref{eq:RefertenceTemperatureD1Q11WithC5-5}: D1Q11. All populations \eqref{eq1:HermiteMu-EDF} with $\rho=1.0$ and weights \eqref{eq:1DmuGeneralizedWeights} are positive with $\mu=0$ and given $\theta = \theta_0$  values provided that: D1Q5: $0 \leq |u| \leq 1.145$; D1Q7: $0 \leq |u| \leq 0.761 $; D1Q9: $0 \leq |u| \leq 0.346$; D1Q11: $0 \leq |u| \leq 1.117$.}\label{tabular:ReferenceTemperatureD1Qnq}
\end{table}

\begin{table}
	\centering
		{  											
		\begin{tabular}[t]{|ccccccccc|}
		\hline
		 & \vline & D1Q5 & \vline &  D1Q7 & \vline &  D1Q9  & \vline &  D1Q11 \\
		\hline
		 & \vline & with & \vline &  with & \vline &  with  & \vline &  with \\
		MB coeff.  & \vline & Eq. \eqref{eq:RefertenceTemperatureD1Q5WithC2-3} & \vline & Eq.  \eqref{eq:RefertenceTemperatureD1Q7WithC3-3} & \vline &  Eq. \eqref{eq:RefertenceTemperatureD1Q9WithC5} & \vline &   Eq. \eqref{eq:RefertenceTemperatureD1Q11WithC5-5}  \\
		\hline
		$Q_3$  & \vline & \textcolor{blue}{\fbox{ \textcolor{black}{ $1$  } }} & \vline & 1  & \vline &  1 & \vline &  1 \\
		$R_2$  & \vline & \textcolor{blue}{\fbox{ \textcolor{black}{ $6$  } }} & \vline &  6 & \vline &  6 & \vline &  6 \\
		$R_4$  & \vline & \textcolor{blue}{\underline{\textcolor{black}{$0$}}}  & \vline & \textcolor{blue}{\fbox{ \textcolor{black}{ $1$  } }}  & \vline & 1  & \vline &  1 \\
		$S_1$  & \vline & \textcolor{blue}{\fbox{ \textcolor{black}{ $15$  } }} & \vline & 15  & \vline & 15  & \vline &  15 \\
		$S_3$  & \vline &  \textcolor{red}{\fbox{\fbox{ \textcolor{blue}{\textcolor{black}{ $6.1257$ }}   }}}  & \vline & \textcolor{blue}{\fbox{ \textcolor{black}{ $10$  } }}  & \vline & 10  & \vline &   10 \\
		$S_5$  & \vline &  \textcolor{blue}{\underline{\textcolor{black}{$0$}}}  & \vline & \textcolor{blue}{\underline{\textcolor{black}{$0$}}}  & \vline & \textcolor{blue}{\fbox{ \textcolor{black}{ $1$  } }}  & \vline &  1 \\
		$V_2$  & \vline &  & \vline & \textcolor{blue}{\fbox{ \textcolor{black}{ $45$  } }}   & \vline & 45  & \vline &  45 \\
		$V_4$  & \vline &  & \vline & \textcolor{red}{\fbox{\fbox{ \textcolor{blue}{\textcolor{black}{ $13.7497$ }}   }}}   & \vline &  \textcolor{blue}{\fbox{ \textcolor{black}{ $15$  } }} & \vline &  15 \\
		$V_6$  & \vline &  & \vline & \textcolor{blue}{\underline{\textcolor{black}{$0$}}}  & \vline & \textcolor{blue}{\underline{\textcolor{black}{$0$}}}   & \vline &  \textcolor{blue}{\fbox{ \textcolor{black}{ $1$  } }} \\
		\hline
		\end{tabular}
		}
			\caption{(Color online) Values of the MB coefficients for different D1Q$n_q$ LB models. The MB coefficients are seen in Eqs. \eqref{eq:AccuracyD1Q51D-3}, \eqref{eq:AccuracyD1Q71D-4}-\eqref{eq:AccuracyD1Q71D-6}. The presented unconditioned matching terms to the MB coefficients are not underlined or single/double boxed. 
			Unconditioned no matching terms are underlined.  
			Single boxed terms are conditioned to $\theta=\theta_0$.  
			Double boxed terms are conditioned to $\theta=\theta_0$, but still they are no matching terms to the MB coefficients. The chosen $\theta_0$ values are seen in table \ref{tabular:ReferenceTemperatureD1Qnq}.}
	\label{tab:MBCoefficientsD1Qnq}
\end{table}

Based on the weights, it is easy to show that the increase of $z$ to a value larger than one leads to a heavy tail in the distribution. The populations $f_i^{\textrm{eq}}$ with $u=0$ and $\rho=1$ for the one-dimensional lattice D1Q$n_q$ models with $n_q=3-11$ and integer $c_k$ values are shown in Fig. \ref{fig:Weights1DmuGeneralizedHigherOrder}. For easy visualization, the  \textit{likely} shapes of the distributions are also drawn by means of interpolation among the discrete weight points of the distributions. From a light-tailed distribution when $z=1$, for the D1Q3 model (\textcolor{black}{\Large{$\circ$} $-$ } (circle-solid)) to heavier tails when $z$ is progressively increased is illustrated in Fig.  \ref{fig:Weights1DmuGeneralizedHigherOrder}. In addition, the increase of the ``temperature'' from $\theta<\theta_0$ to $\theta_0$ leads to smaller kurtosis as it is seen for the cases D1Q5 (from \textcolor{black}{$\square$ $\cdot$ $\cdot$} (squared-dotted), \textcolor{black}{\Large{$\vartriangle$} -. } (triangle up-dashdot) to \textcolor{blue}{\Large{$\triangledown$} $-$} (triangle down-solid)) and D1Q11 (from \textcolor{cyan}{\Large{$\triangleright$} - -} (triangle right-dashed) to \textcolor{magenta}{\Large{$\triangledown$} - -} (triangle down-dashed)). In these two cases, the peakedness of the distributions are significantly affected. On the other hand, the D1Q7 and D1Q9 lattice models do not show such properties when $\theta=\theta_0$, c.f. \textcolor{black}{\Large{$\triangleleft$} $-$} (triangle left-solid) and \textcolor{red}{$\times$ $\cdot$ $\cdot$} (cross-dotted) in Fig. \ref{fig:Weights1DmuGeneralizedHigherOrder}. Coincidentally, these two particular D1Q5 and D1Q11 models have the largest $\theta_0$ values, as seen from the outlined values of $\theta_0 \approx 0.33, \textbf{1.63}, 0.69, 0.75, \textbf{2.12}$ for $n_q= 3, \textbf{5}, 7, 9, \textbf{11}$ respectively. That is, fatter tails are obtained with the increase of the $\theta$ value, as observed in Fig. \ref{fig:Weights1DmuGeneralizedHigherOrder}.

\begin{figure}[!htbp]
  \begin{center}
\includegraphics[angle=0, width=1.0\textwidth]{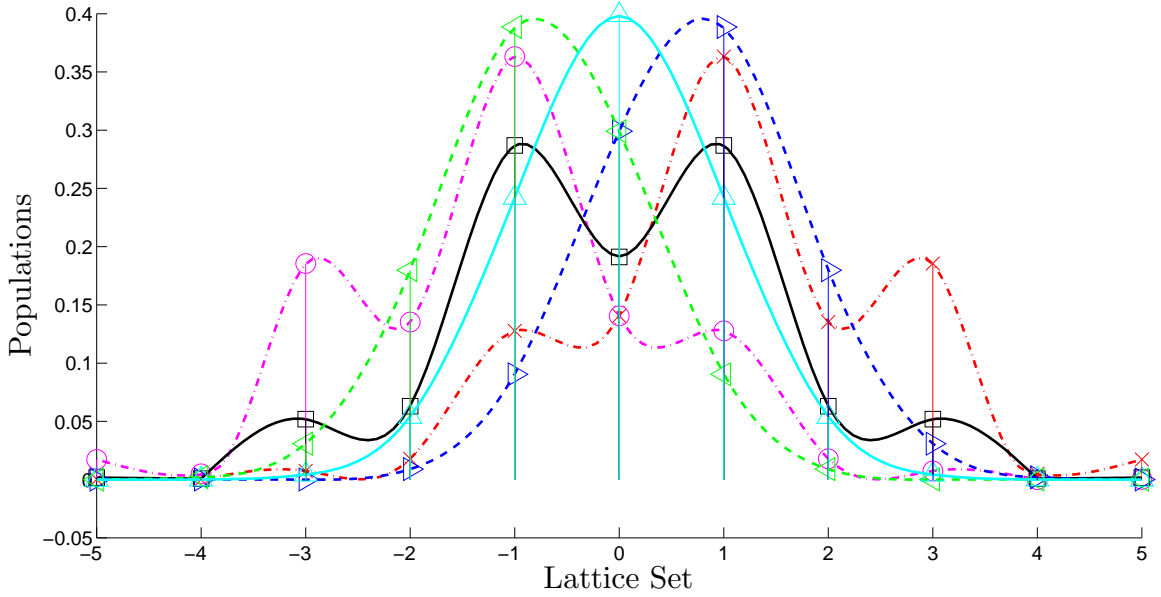}
    \caption{(Color online) Populations obtained from Eqs. \eqref{eq:1DmuGeneralizedWeights} and \eqref{eq1:HermiteMu-EDF}  with $\mu=0$, $\rho=1$, and the \textit{likely} shapes of their distributions for the D1Q11 model with integers lattice velocity set $\boldsymbol{c} = \{  -5, -4, -3, -2, -1, 0, 1, 2, 3, 4, 5 \}$. 
    Symbols: 
     \textcolor{cyan}{\Large{$\vartriangle$} $-$} (triangle-solid): with $\theta=1.0$ and $u=0$;  
     \textcolor{blue}{\Large{$\triangleright$} - -} (triangle right-dashed): with $\theta=1.0$ and $u=0.74$;  
     \textcolor{green}{\Large{$\triangleleft$} - -} (triangle left-dashed): with $\theta=1.0$ and $u=-0.74$;  
    \textcolor{black}{$\square$ $-$} (squared-solid): with Eq. \eqref{eq:RefertenceTemperatureD1Q11WithC5-5} and $u=0$; 
    \textcolor{red}{$\times$ - .} (cross-dashdot): with Eq. \eqref{eq:RefertenceTemperatureD1Q11WithC5-5} and $u=1.117$; 	
    \textcolor{magenta}{\Large{$\circ$} - .} (circle-dashdot): with Eq. \eqref{eq:RefertenceTemperatureD1Q11WithC5-5} and $u=-1.117$.}
    \label{fig:PopulationsAndWeightsforD1Q11CasesHigherOrder}
  \end{center}
\end{figure}

The shape of the distribution is also altered due to the flow velocity. 
Recall that the populations $f_i^{\textrm{eq}}$ (c.f. Eqs. \eqref{eq:1DmuGeneralizedWeights} and \eqref{eq1:HermiteMu-EDF} with $\mu=0$) are $\rho$-scaled velocity perturbations on the weights. An example is depicted in Fig. \ref{fig:PopulationsAndWeightsforD1Q11CasesHigherOrder}  for the lattice set D1Q11. Here, skewness is affected with the increase of the velocity.  With $\theta=1.0$ and maximum (minimum) possible lattice velocity $u= 0.74$ ($u= -0.74$), the distribution is skewed to the left (right). Negative populations are obtained with further increase of the lattice velocity  $|u|$.  For instance, with populations denoted as $\{  f_{-5}, f_{-4}, f_{-3}, f_{-2}, f_{-1}, f_0, f_1, f_2, f_3, f_4, f_5 \}$ on the lattice set and with $u=0.75$ ($u=-0.75$) yields $f_{-3} < 0$ ($f_{3} < 0$); with  $u=1.1$ ($u=-1.1$) yields $f_{-3} < 0$ and $f_{-2} < 0$ ($f_{3} < 0$ and $f_{2} < 0$); with  $u=1.35$ ($u=-1.35$) yields $f_{-3} < 0$, $f_{-2} < 0$ and $f_{0} < 0$ ($f_{3} < 0$, $f_{2} < 0$ and $f_{0} < 0$); and so on. With the reference value of Eq. \eqref{eq:RefertenceTemperatureD1Q11WithC5-5}, the MB $(z+1)$-moment is now  guaranteed with $z=5$, leading to a new maximum (minimum) possible lattice velocity of $u= 1.117$ ($u= -1.117$), and the distribution is skewed to the left (right). In this case, negative populations start to show up with   $u=1.12$ ($u=-1.12$) and it yields $f_{-4} < 0$ ($f_{4} < 0$); with  $u=1.5$ ($u=-1.5$) yields $f_{-4} < 0$ and $f_{-3} < 0$ ($f_{4} < 0$ and $f_{3} < 0$); with  $u=2$ ($u=-2$) yields  $f_{-4} < 0$, $f_{-3} < 0$ and $f_{0} < 0$ ($f_{4} < 0$, $f_{3} < 0$ and $f_{0} < 0$); and so on. That is, the first negative population is obtained where the tail is longer. Hence, the first source of instability at a fixed $\theta$ value shows up on the part from which the distribution is skewed to. Upon fulfilling some high-order hydrodynamic moments, the presence of ``thermal'' tails in the distributions allows capturing high velocity particles. 

How large the $z$ value must be depends on the needed MB moments to be fulfilled, which in turns is determined by the particular case to simulate. In general, flows with high velocities and temperature values require large $z$ values. The increase of $z$ and the (allowed) temperature values lead to longer and fatter tails, as already mentioned. Normally, the (asymptotic) extremes are adopted as a starting point to study models, from which their behaviors in between these two sides are later considered.  Most of the LB works found in the literature are based on the low order construction, i.e. $z=1$. The other extreme, $z  \rightarrow \infty $, is now considered for the present construction. 
The aim here is not to go deep into theoretical descriptions, which can derail this work from the LB method, but to have, at least, a general idea about some possible/expected properties of the distribution for very large $z$ values. Then, the study can be possible linked to an existing theory, from which further work can be conducted elsewhere. 
The study of the tail distribution involves the use of the cumulative distribution function (CDF), $F(c_n) = \sum_{c_k=-z}^{c_n \leq z} W_{c_k}$, and the complementary cumulative distribution function (CCDF),  $\overline{F}(c_n) = 1- F(c_n)$, where $c_n$ are integers numbers in the range of $[ -z, -(z-1), \dots, -1, 0, 1, \dots, (z-1), z ]$. 
The CDF can be entirely written in terms of the off-centered lattice cell weights only, e.g. Eq. \eqref{eq:1DmuGeneralizedWeightsk}, i.e. 
\begin{eqnarray} \nonumber
F(c_n) &=& (-1)^{H_1(c_n)} \sum_{c_k=-z}^{c_n \leq -1}  W_{c_k} + H_1(c_n) \ \textrm{sgn}(c_n) \sum_{c_k=c_n \geq 1}^{c_n \leq z} W_{c_k} + H_1(c_n), \\ \label{eq:CumulativeDistributionFunction}
\end{eqnarray} 
where $H_a(c_n)$ is the Heaviside step function 
\begin{eqnarray} \label{eq:HeavisideFunction}
H_a(c_n) &=&  
\left\{ 
\begin{array}{lll}
0, & c_n < 0, \\
a, & c_n = 0, \\
1, & c_n > 0,
\end{array} 
\right. 
\end{eqnarray} 
so that $H_a(0)=a$ is also valid and  $\textrm{sgn}(c_n) = 2 H_{1/2}(c_n) -1$. 

\begin{figure}[!htbp]
  \begin{center}
\includegraphics[angle=0, width=1.0\textwidth]{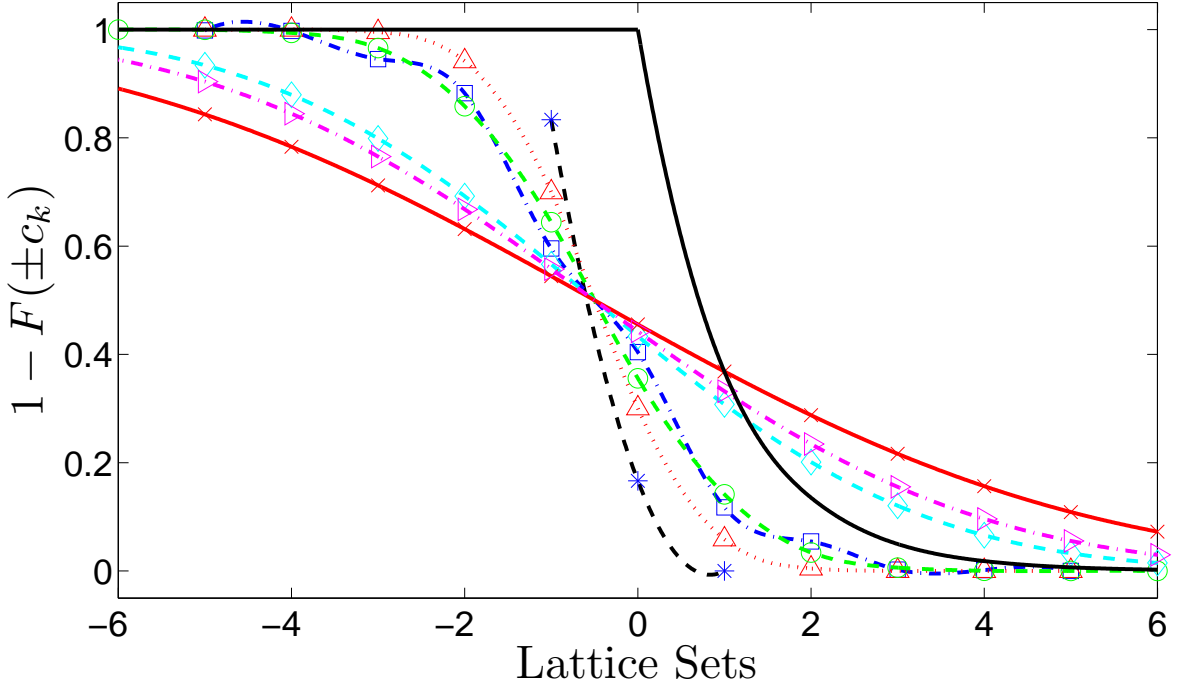}
    \caption{(Color online) Complementary cumulative distribution function (CCDF) of the weights (populations with $u=0$, $\rho=1$) obtained from Eqs. \eqref{eq:1DmuGeneralizedWeights} with $\mu=0$, and their \textit{likely} shapes for the D1Q$n_q$ models with integers lattice velocity set $\boldsymbol{c} = \{  -z, -(z-1), \dots, -1, 0, 1, \dots, z-1, z \}$ and $n_q=3, 11, 13, 81$ and $201$, which correspond to $z =1, 5, 6, 40$ and $100$ respectively. 
    Symbols: 
     \textcolor{blue}{\Large{$\ast$} } - - (asterisk-dashed): D1Q3 with $\theta = \theta_0 = 1/3$;  
     \textcolor{red}{\Large{$\vartriangle$} . .} (triangle-dotted): D1Q11 with $\theta=1.0$;  
    \textcolor{blue}{$\square$ - .} (squared-dashdot): DQ11 with Eq. \eqref{eq:RefertenceTemperatureD1Q11WithC5-5}; 
    \textcolor{green}{\Large{$\circ$} - -} (circle-dashed): DQ13 with $\theta = 2.0$;
    \textcolor{cyan}{$\diamondsuit$ - -} (diamond-dashed): D1Q81 with $\theta=9.0$;  
    \textcolor{magenta}{\Large{$\triangleright$} - .} (triangle right-dashdot): D1Q201 with $\theta=12.0$;  
    \textcolor{red}{$\times$ $-$} (cross-solid): D1Q201 with $\theta=20.0$;  
    Solid: CCDF of the exponential distribution. The weights are positive within machine precision.}
    \label{fig:SubExponential}
  \end{center}
\end{figure}

Some (zoomed) CCDF for D1Q$n_q$ models with $n_q=3, 11, 13, 81$ and $201$, which correspond to $z =1, 5, 6, 40$ and $100$ respectively, are plotted in Fig. \ref{fig:SubExponential} for difference $\theta$ values and $\mu=0$. 
In general, the decay to zero of the CCDF becomes slower when $z$ is increased, as seen in Fig. \ref{fig:SubExponential}, i.e. from a rather upright CCDF for D1Q3 (\textcolor{blue}{\Large{$\ast$} } - - (asterisk-dashed)) to a more horizontal CCDF for D1Q201 (\textcolor{red}{$\times$ $-$} (cross-solid)). Distributions with the observed characteristics in Fig.  \ref{fig:SubExponential} can be long-tailed and subexponentials.


\textit{Some} basic properties of subexponential distributions (at infinity) are \cite{EmbrechtsKluppelbergMikosch1997Book}: 
\begin{subequations} \label{eq:Long-TailedNecessaryConditionSubexponentialsubexponential}
\begin{eqnarray}  \label{eq:Long-Tailed}
(1) &\quad& \lim_{c_n \rightarrow \infty} \frac{\overline{F}(c_n-y)}{\overline{F}(c_n)} = 1, \forall \ 0 < y < \infty, \\ \label{eq:subexponential}
(2) &\quad& \lim_{c_n \rightarrow \infty} e^{s c_n} \overline{F}(c_n) = \infty,  \forall s>0. 
\end{eqnarray} 
\end{subequations}
Rigorous proofs of \eqref{eq:Long-TailedNecessaryConditionSubexponentialsubexponential} for more general functions are found in \cite{EmbrechtsKluppelbergMikosch1997Book} and are not repeated in this work. Here, sketched proofs and examples are outlined instead, as an attempt to present the subject more accessible and intuitive to LB practitioners. The interpretation of the relations \eqref{eq:Long-TailedNecessaryConditionSubexponentialsubexponential} are then used as links to show some trends and properties of the distribution \eqref{eq:1DmuGeneralizedWeights} throughout examples (for some finite $z$ values). 


Recall that the present construction allows integer lattice velocities and thus $y, c_n$ and $z$ are considered integers, i.e. $1 \leq y < c_n \leq z$. The trivial solution $y=0$ is obviously excluded. When $c_n > 0$ the $H_1(c_n)=1$ and $\textrm{sgn}(c_n)=1$ and then, the CCDF of distribution becomes 
\begin{subequations} \label{eq:CCDFPartialResults}
\begin{eqnarray}  \nonumber
\overline{F}(c_n) &=& 1- F(c_n) =\{ \textrm{ with Eq. \eqref{eq:CumulativeDistributionFunction} } \} \\ \nonumber
&=& 1 - \Bigg(  - \sum_{c_k=-z}^{-1} W_{c_k} +  \sum_{c_k= 1}^{c_n \leq z} W_{c_k} + 1 \Bigg) \\ \label{eq:CCDFPartialResultsA}
&=& \sum_{c_k=-z}^{-1} W_{c_k} - \sum_{c_k= 1}^{c_n \leq z} W_{c_k} \\ \nonumber
&=& \{ \textrm{ with } W_{-c_k} = W_{c_k}  \} \\ \label{eq:CCDFPartialResultsB} 
&=& \underbrace{W_{z} + W_{z-1} + \cdots + W_{c_n+1}}_{I},  
\end{eqnarray} 
\end{subequations}
where the result in \eqref{eq:CCDFPartialResults}, e.g. the term $I$ which is $\sum_{c_k=c_n+1}^{z} W_{c_k}$, is zero when $c_n=z$ by definition. That is, the number of summands $ns$ in $\sum_{c_k=c_n+1}^{z} W_{c_k}$ is so that $ns + c_n = z$. 
Similarly, for $\overline{F}(c_n-y) = \sum_{c_k=c_n+1-y}^{z} W_{c_k}$ when $c_n > 0$. 
These weights values in $\overline{F}(c_n)$, \eqref{eq:CCDFPartialResults}, and in $\overline{F}(c_n-y)$ when $0<y<c_n \gg 0$  correspond to those located at the extreme of the tail. These extreme weights get closer to zero when $z$ is increased, as they are presented (in the last column) in table \ref{tab:WeightsValuesExtremeOfTail}. 
Hence, for a very large $c_n \leq z$ the expression \eqref{eq:Long-Tailed} becomes a ratio between zeros. The l'H\^{o}pital's rule can be applied to evaluate this limit. The weights \eqref{eq:1DmuGeneralizedWeights} can be expressed in terms of elementary symmetric polynomials, c.f. Eqs. \eqref{eq:MaxTemp}, and the derivatives of such polynomials  \cite{RahmanSchmeisserBook} are out of the scope of this work. 

The expression \eqref{eq:Long-Tailed} is a property of slowly varying functions (at infinity). As already noted in this work, the observed trend in Fig. \ref{fig:SubExponential} is that the CCDF varies slower to zero when $z$ is increased. For instance,  from a fast varying CCDF for D1Q3 (\textcolor{blue}{\Large{$\ast$} } - - (asterisk-dashed)), 
to a slower CCDF for D1Q11 with $\theta=1.0$ (\textcolor{red}{\Large{$\vartriangle$} . .} (triangle-dotted)), 
which is further changed with Eq. \eqref{eq:RefertenceTemperatureD1Q11WithC5-5} (\textcolor{blue}{$\square$ - .} (squared-dashdot)).   
The variation is shown in Fig. \ref{fig:SubExponential} progressively, for D1Q13 (\textcolor{green}{\Large{$\circ$} - -} (circle-dashed)) with $\theta = 2.0$, to D1Q81 with $\theta=9.0$ (\textcolor{cyan}{$\diamondsuit$ - -} (diamond-dashed)), to 
D1Q201 with $\theta=12.0$ (\textcolor{magenta}{\Large{$\triangleright$} - .} (triangle right-dashdot)),  
which is further changed with $\theta=20.0$ (\textcolor{red}{$\times$ $-$} (cross-solid)).

\begin{table}
	\centering
		{  											
		\begin{tabular}[t]{|ccccccc|}
		\hline
		 Model & \vline & $\theta$ & \vline & Fig. \ref{fig:SubExponential} & \vline &  Eq. \eqref{eq:1DmuGeneralizedWeightsk}  \\ 
		\hline
		 D1Q3 & \vline & $\theta_0=1/3$ & \vline & \textcolor{blue}{\Large{$\ast$} } - - & \vline &  $W_{z=1}=1/6$ \\
		  (z=1) & \vline &   & \vline & (asterisk-dashed) & \vline &   \\
		\hline
		 D1Q11 & \vline & $1.0$ & \vline & \textcolor{red}{\Large{$\vartriangle$} . .} & \vline & $W_{z-1=4} \approx 1.3 \times 10^{-4}$, \\
		(z=5)  & \vline & & \vline & (triangle-dotted) & \vline & $W_{z=5} \approx 1.6 \times 10^{-6}$   \\
		\hline
		 D1Q11 & \vline & Eq. \eqref{eq:RefertenceTemperatureD1Q11WithC5-5} & \vline & \textcolor{blue}{$\square$ - .}  & \vline & $W_{z-1=4} \approx 8.2 \times 10^{-4}$, \\
		(z=5) & \vline & & \vline & (squared-dashdot) & \vline & $W_{z=5} \approx 1.6 \times 10^{-3}$   \\
		\hline
		 D1Q13 & \vline & $2.0$ & \vline & \textcolor{green}{\Large{$\circ$} - -}   & \vline & $W_{z-1=5} \approx 3.9 \times 10^{-4}$, \\
		(z=6) & \vline & & \vline & (circle-dashed) & \vline & $W_{z=6} \approx 5.7 \times 10^{-5}$   \\
		\hline
		 D1Q81 & \vline & $9.0$ & \vline & \textcolor{cyan}{$\diamondsuit$ - -}    & \vline & $W_{z=30-40}= a \times 10^b$, \\
		(z=40) & \vline & & \vline & (diamond-dashed) & \vline & $a \sim$ one digit, \\
		& \vline & & \vline &  & \vline & $b \sim$ $[ -25 \cdots -40]$  \\
		\hline
		 D1Q201 & \vline & $12.0$, & \vline & \textcolor{magenta}{\Large{$\triangleright$} - .}    & \vline & $W_{z=90-100}= a \times 10^b$, \\
		(z=100) & \vline &  & \vline & (triangle right-dashdot) & \vline & $a \sim$ one digit, \\
		& \vline & $20.0$ & \vline & \textcolor{red}{$\times$ $-$} & \vline & $b \sim$ $[ -80 \cdots -100]$  \\
		& \vline &  & \vline & (cross-solid) & \vline &   \\
		\hline

		\end{tabular}
		}
			\caption{(Color online) Values of the weights located at the extreme of the tail for D1Q$n_q$ model, where $n_q=3, 11, 13, 81, 2001$, c.f. Fig. \ref{fig:SubExponential}.  The symbol $\sim$ represents ``of the order of''.}
	\label{tab:WeightsValuesExtremeOfTail}
\end{table}

Eq. \eqref{eq:Long-Tailed} can be easily rewritten as $\lim_{c_n \rightarrow \infty} \overline{F}(s\ c_n)/\overline{F}(c_n) \rightarrow 1$ with $s>0$. With $x=e^{c_n}$, Eq. \eqref{eq:subexponential} becomes $x^s \ \overline{F}(\textrm{ln}(x))$. 
By definition $\overline{F}(\textrm{ln}(x)) \rightarrow 0$ when $x \rightarrow \infty$ as a result of $x=e^{c_n}$ and $c_n \rightarrow \infty$. On the other hand, $x^s \rightarrow \infty$ with $s > 0$ for similar reasons. The exponential CDF is 
\begin{eqnarray} \label{eq:CDFExponentialDistribution}
F_{\textrm{exp}}(c_n) &=&  
\left\{ 
\begin{array}{ll}
1 - \textrm{exp}(-s c_n), & c_n \geq 0, \\
0, & c_n <0. 
\end{array} 
\right. 
\end{eqnarray} 
The exponential CCDF with $s=1$ is plotted in Fig. \ref{fig:SubExponential} as a solid line-curve. From 
the positive side of the lattice sets, i.e. $c_i \geq 0$ in Fig. \ref{fig:SubExponential} is observed that the decay of the CCDF for the D1Q81 and D1Q201 models are slower than the corresponding exponential CCDF. Thereby the name of ``subexponential''. Similar studies can be done for $s > 0$, $s \neq 1$. 



Although it can be argued that some (physical) phenomena can be described or detected by models with long-tailed subexponential  distributions, which in turn can be linked to extreme value theory \cite{EmbrechtsKluppelbergMikosch1997Book}, further analysis is required. In addition, $n$-modal (i.e. with $n$-peaks) distributions are also observed in Figs.  
\ref{fig:Weights1DmuGeneralizedHigherOrder} and \ref{fig:PopulationsAndWeightsforD1Q11CasesHigherOrder}. Such studies deserve separate works elsewhere.

A feasible high-order LB D1Q5 model with a lattice set $\{ c_0, \pm c_1, \pm c_2 \}$ $= \{ 0, \pm 1, \pm 2 \} $ has received a  deserved attention in the literature \cite{QianZhou1998}, \cite{Dellar2005}, \cite{ChikatamarlaKarlin1D2006}, \cite{ChikatamarlaKarlin2009} because it would be a good model for the Navier-Stokes equation  \eqref{eq:IsothermalNavier-Stokes} with the shortest \textit{on}-Cartesian lattice set (Fig. \ref{fig:Cartesian} d)). However, in previous (isothermal) models, this one-dimensional five velocity model $\{ 0, \pm 1, \pm 2 \} $, \cite{QianZhou1998}, proves intrinsic unstable \cite{Dellar2005}, due to complex reference ``temperature'' $\theta_0$ value    \cite{ChikatamarlaKarlin1D2006}. 
Many causes have been attributed in order to answer the reason of such instabilities. 
The following statement is found in \cite{ChikatamarlaKarlin2009}: 
``In some of the earlier studies, the pattern of instability of the  $\{0, \pm 1, \pm 2 \}$ lattice, was attributed to the lattice Boltzmann scheme itself  \cite{McNamaraGarciaAlder1997}, or to the advection part of the LB scheme   \cite{SiebertHegelePhilippi2008}, \cite{Dellar2005}, or to the collision of the LB scheme  \cite{BrownleeGorbanLevesley2007}, or to insufficient isotropy \cite{NieShanChenEuroPHys2008}''.  
Although the problem is identified in the afore-cited references, different strategies are adopted to tackle the issue, 
not necessarily following the on-Cartesian approach. 
In \cite{ChikatamarlaKarlin1D2006} (which follows the on-Cartesian approach), the blame was put on the lattice, e.g. for the $ \{ 0, \pm 1, \pm 2 \} $ lattice case. 
It is shown later in this work that the proposed construction in this paper is capable to have real-valued  reference "temperature" $\theta_0$  with the shortest on-Cartesian lattice sets. 

Because of the high-order LB construction in \cite{ChikatamarlaKarlin1D2006}, \cite{ChikatamarlaKarlin2009} is reported to be limited up to DdQ$(9)^d$ (more about this below), no likely trend has been described in the literature about which lattice patterns have problems when $c_k$ with $k= 0, 1, 2, \dots$, i.e. consecutive integers. For the moment $\sum f_i^{\textrm{eq}} c_{\alpha,i}^{(z+1)}$, a \textit{likely} trend is observed in the present construction in which the D1Q$n_q$ models with $n_q =3, 5, 9$ have complex-valued $\theta_0$ when $c_k = k$, where $k=0, 1, \dots, z$, while the models with $n_q = 5+2, 9+2= 7, 11$ and $c_k = k$ have no complex-valued $\theta_0$. The problem with the former models can be avoided, for instance, by using $c_k = k$ with $k = 0, 1, \dots, (z-1)$, while $c_{k} \geq (z+1)$ when $k=z$. An example: $\{ c_0, \pm c_1, \pm c_2, \pm c_3, \pm c_4 \}$ $= \{ 0, \pm 1, \pm 2, \pm 3, \pm 5 \} $ for the D1Q9. 
These results have been algebraically tested up to $n_q=11$ in a general form. Cases with $n_q>11$, under the same algebraic conditions, are computational demanding, which are out of the scope of this work. 

It should be pointed out that every complex-valued $\theta_0$ leads to complex-valued weights and thereby to complex-valued  populations, which is nonsense. From probability theory, populations are nonnegative real-valued. Some combinations of non-consecutive $c_k$, e.g. $\{0, \pm 1, \pm 3, \pm \geq 5 \} $ for the D1Q7 model, can still give a complex valued $\theta_0$. Any $\theta_0$ value, which lead to negative populations, will contribute to instability, no matter whether they are real valued or not. Therefore, the calculated and presented $\theta_0$ values in this work (c.f. tables \ref{tabular:ReferenceTemperatureD1Qnq} and \ref{tabular:GeneralizedReferenceTemperatureD1Qnq} ) are confirmed to give positive weights and populations within a given flow velocity range (in lattice units).

All the presented relations in this work have been obtained using the Hermite construction approach. Any belief that the aforementioned results are merely obtained from the so called ``entropic'' construction \cite{ChikatamarlaKarlin2009} is discarded from the current results. For instance, Eq. \eqref{eq:ReferenceTemperatureD1Q5} is found for the same one-dimensional lattice with $n_q=5$ in \cite{ChikatamarlaKarlinProductForm2006} (c.f. Eqs. (10)-(11) therein). Furthermore, the values from Eqs. \eqref{eq:RefertenceTemperatureD1Q7WithC3-3} and \eqref{eq:RefertenceTemperatureD1Q9WithC4-5-8} are the same as those obtained in \cite{ChikatamarlaKarlin2009} (c.f. relations (9), (C3) and (D3), (D5) therein respectively), although presented in this work with higher accuracy. Finally, the isothermal on-Cartesian lattice weights found in \cite{ChikatamarlaKarlin2009} 
can be also obtained from the Eq. \eqref{eq:1DmuGeneralizedWeights} (with $\mu=0$) together with their respective $\theta_0$ in Eq. \eqref{eq:ReferenceTemperatureD1Qnq}. 
These similarities on isothermal weights and $\theta_0$ values should not come as a surprise, taking into account how the weights and $\theta_0$ values can be obtained in the ELB construction to match MB moments (c.f. appendix A in \cite{KarlinElementsLBMII}). Similar outputs can be obtained when equivalent moments are forced to be matched. 




Having mentioned some similarities between the actual thermal Hermite-based construction and the isothermal ``entropic'' one (reviewed) in \cite{ChikatamarlaKarlin2009}, their differences are substantial and cannot be overemphasized. 
Unlike the Hermite-based construction, the ELB method relies on macroscopic equations and its physical extension beyond these descriptions is based on adding lattice velocities, at the expense of the presence of (high) powered spurious velocity terms, c.f. Eqs. (8), (10) and (D6) in \cite{ChikatamarlaKarlin2009} and appendix. There are no spurious velocity terms in the relations obtained from the current construction, c.f. \eqref{eq:AccuracyD1Q51D} and \eqref{eq:AccuracyD1Q71D}. 
The one (summarized) in \cite{ChikatamarlaKarlin2009} is based on an isothermal construction, i.e. mass and momentum density, and the gained $\theta$ (through constraints) is used as a manipulation tool. Hence, from the pressure tensor and beyond, the results are isothermal and spurious terms are obtained. Even if the Eq. \eqref{eq:Complete-product-form} in appendix is used to construct a high-order ELB model, 
the procedure is still limited to mass, momentum density and (the trace of) the pressure tensor. Manipulations are then needed to guarantee MB moments beyond those descriptions, say by sacrificing $\theta$, and spurious terms show up \textit{from} the energy flux and beyond. The existence of spurious velocity terms limits any approach to $u<1$ so that they can be neglected. Note that in the current construction, the D1Q11 model with fixed $\theta=\theta_0$ has maximum possible velocity of $u=1.117$. The approach in \cite{ChikatamarlaKarlin2009} is outlined for ``\textit{all} possible discrete velocity sets, in one dimension'', and subsequently presented up to nine-velocity set. One can argue that higher order ELB models can be constructed, but the flow velocity has to be limited due to the existence of spurious velocity terms. Although the relations \eqref{eq:1DmuGeneralizedWeights} and   \eqref{eq:MaxTemp} have been  algebraically obtained up to the lattice D1Q13 model while \eqref{eq1:HermiteMu-EDF} up to the D1Q11 in a general form, no mathematical restrictions are imposed in this 
work, no spurious velocity terms arise for the first MB $(z+1)$-moments and the value of $z$ can be theoretically larger than that. For instance, some weight values computed for $n_q>13$ are plotted in Fig. \ref{fig:SubExponential}. 
There are no rational approximations for the  $u^{z+2}$ velocity term of the MB coefficient belonging to the MB $(z+2)$ moment in the current construction, contrary to what is found in \cite{ChikatamarlaKarlin1D2006} for the lattice set D1Q5 model. In the present work, these  MB coefficients related to the $u^{z+2}$  terms are zero unconditionally, c.f. $R_4=0$, $S_5=0$ and $V_6=0$ in table \ref{tab:MBCoefficientsD1Qnq} for D1Q5, D1Q7 and D1Q9 models respectively. 

Note that the use of the weights \eqref{eq:1DmuGeneralizedWeights} with \eqref{eq1:HermiteMu-EDF} and $\mu=0$ leads to thermally matched MB $z$-moments exactly, mathematically speaking, with \textit{integer} lattice velocities $c_k$ for the D1Q$n_q$ high-order models, as already mentioned above. Hence, no interpolations or approximations are theoretically needed to coincide with the Cartesian grid nodes, c.f. Fig. \ref{fig:Cartesian}. The present thermal construction is more general and accurate than the isothermal low Mach number approach (summarized) in \cite{ChikatamarlaKarlin2009}.  

Similarly, there exist a link between the computed weights from Eq. \eqref{eq:1DmuGeneralizedWeights} with $\mu=0$ and those found in the literature by other authors, e.g. in \cite{ShanYuanChen2006}. For instance, the roots of the fifth-order Hermite polynomial $H_5^{(0)} (x/\sqrt{2})=0$ are $0, \pm \sqrt{ 5 - \sqrt{2 \cdot  5} }$, $\pm \sqrt{ 5 + \sqrt{2 \cdot  5} }$. With $z=2$, $c_1=\sqrt{ 5 - \sqrt{2 \cdot  5} }$ and $c_2=\sqrt{ 5 + \sqrt{2 \cdot  5} }$ in Eqs. \eqref{eq:ReferenceTemperatureD1Q5} and \eqref{eq:1DmuGeneralizedWeights} lead to $\theta_0=1$ and 
\begin{subequations} \label{eq:athermalWeightsD1Q5HermiteRoots}
\begin{eqnarray} 
		W_0 &=& \frac{8}{15}, \\
		W_{1,2} &=& \frac{7 + 2 \sqrt{2 \cdot 5}}{60}, \\
		W_{3,4} &=& \frac{7 - 2 \sqrt{2 \cdot 5}}{60},
\end{eqnarray} 
\end{subequations}
respectively. Hence, in this context, the current construction 
can be reduced to the particular case of athermal weights in \cite{ShanYuanChen2006}, where the results in Eqs. \eqref{eq:athermalWeightsD1Q5HermiteRoots} are found in table 1 therein. Note that with $c_1=\sqrt{ 5 - \sqrt{2 \cdot  5} }$ and $c_2=\sqrt{ 5 + \sqrt{2 \cdot  5} }$ in Eq.  \eqref{eq:MBcoefficientForD1Q5-S3} the $S_3=10 =S_3^{\textrm{MB}}$, while the rest of the MB coefficients remain the same as they are presented in table \ref{tab:MBCoefficientsD1Qnq} for the D1Q5 model. 
That is, only the MB coefficient related to the $u^{z+1}$ velocity term belonging to the MB $(z+3)$ moment is improved when the the aforementioned non-integers lattice $c_k$ are implemented. This, at the price of using, for example interpolations due to the presence of Cartesian-lattice mismatches,  c.f. Fig. \ref{fig:Cartesian}. Thus the computational cost is increased and the main LB idea is not guaranteed. Hence, the choice of \textit{on}-Cartesian integer lattice velocity $c_k$ seems more appealing.

\begin{table}
	\centering
		{\scriptsize  											
		\begin{tabular}[t]{|ccccccccc|}
		\hline
		Coeff. & \vline & D1Q5 & \vline &  D1Q7 & \vline &  D1Q9  & \vline &  D1Q11 \\
		\hline
		$Q_3$  & \vline & \textcolor{blue}{\fbox{ \textcolor{black}{ $1$  } }} & \vline & 1  & \vline &  1 & \vline &  1 \\
		$R_2$  & \vline & \textcolor{blue}{\fbox{ \textcolor{black}{ $2 (3+2 \mu)$  } }} & \vline &  $2 (3+2 \mu)$ & \vline &  $2 (3+2 \mu)$ & \vline &  $2 (3+2 \mu)$ \\
		$R_4$  & \vline & \textcolor{blue}{\underline{\textcolor{black}{$0$}}}  & \vline & \textcolor{blue}{\fbox{ \textcolor{black}{ $1$  } }}  & \vline & 1  & \vline &  1 \\
		$S_1$  & \vline & \textcolor{blue}{\fbox{ \textcolor{black}{ $(3 +  2 \mu)(5 + 2 \mu)$  } }} & \vline & $(3+2 \mu) (5+2 \mu)$  & \vline & $(3 +  2 \mu)(5 + 2 \mu)$  & \vline &  $(3 +  2 \mu)(5 + 2 \mu)$  \\
		$S_3$  & \vline &  \textcolor{red}{\fbox{\fbox{ \textcolor{blue}{\textcolor{black}{ $\cdots$ }}   }}}  & \vline & \textcolor{blue}{\fbox{ \textcolor{black}{ $2 (5+2 \mu)$  } }}  & \vline & $2 (5+2 \mu)$  & \vline &   $2 (5+2 \mu)$ \\
		$S_5$  & \vline &  \textcolor{blue}{\underline{\textcolor{black}{$0$}}}  & \vline & \textcolor{blue}{\underline{\textcolor{black}{$0$}}}  & \vline & \textcolor{blue}{\fbox{ \textcolor{black}{ $1$  } }}  & \vline &  1 \\
		$V_2$  & \vline &  & \vline & \textcolor{blue}{\fbox{ \textcolor{black}{ $3 (3+2 \mu) (5+2 \mu)$  } }}   & \vline & $3 (3+2 \mu) (5+2 \mu)$   & \vline &  $3 (3+2 \mu) (5+2 \mu)$  \\
		$V_4$  & \vline &  & \vline & \textcolor{red}{\fbox{\fbox{ \textcolor{blue}{\textcolor{black}{ $\cdots$ }}   }}}   & \vline &  \textcolor{blue}{\fbox{ \textcolor{black}{ $3 (5+2 \mu)$  } }} & \vline &  $3 (5+2 \mu)$  \\
		$V_6$  & \vline &  & \vline & \textcolor{blue}{\underline{\textcolor{black}{$0$}}}  & \vline & \textcolor{blue}{\underline{\textcolor{black}{$0$}}}   & \vline &  \textcolor{blue}{\fbox{ \textcolor{black}{ $1$  } }} \\
		\hline
		\end{tabular} }
			\caption{(Color online) Values of the $\mathcal{M}$ coefficients for different D1Q$n_q$ LB models. The coefficients are seen in Eqs. \eqref{eq:AccuracyD1Q51D-3}, \eqref{eq:AccuracyD1Q71D-4}-\eqref{eq:AccuracyD1Q71D-6}. The presented unconditioned matching terms to the coefficients are not underlined or single/double boxed. 
			Unconditioned no matching terms are underlined.  
			Single boxed terms are conditioned to $\theta=\theta_0$, where $\theta_0 =$ function$(c_i, \mu)$.  
			Double boxed terms are conditioned to $\theta=\theta_0$, but still they are no matching terms to the coefficients.  $Q_1=(3 + 2 \mu)$ and $R_0=(1 + 2 \mu)(3 + 2 \mu)$ in Eqs. \eqref{eq:AccuracyD1Q51D-3} and \eqref{eq:AccuracyD1Q51D-4}  respectively. $V_0=(1 + 2 \mu)(3 + 2 \mu)(5 + 2 \mu)$ in Eq. \eqref{eq:AccuracyD1Q71D-6}.} 
	\label{tab:GeneralizedCoefficientsD1QnqMU}
\end{table}

The implementation of the classical Hermite polynomial shows that for certain lattice sets (c.f.  results \eqref{eq:RefertenceTemperatureD1Q5WithC2-2} and \eqref{eq:RefertenceTemperatureD1Q9WithC4} in table \ref{tabular:ReferenceTemperatureD1Qnq}) the hydrodynamic $(z + 1)$-moments are \textit{not} matched with the shortest on-Cartesian lattice sets with some fixed real-valued  $\theta$. Hence another construction is needed to accomplish it, whilst preserving the the \textit{on}-Cartesian and non-fixed $\theta$ value properties for the other hydrodynamic $z$-moments.  
At the end of section \ref{section:GeneralizedHermite}, the pressure tensor $(P)$ and energy flux ($Q$), computed from a thermal high-order $\mu$-generalized Hermite-based LB construction (for D1Q$n_q$, $n_q \geq 7$), are used to obtain the lattice pressure and kinematic viscosity. The analysis previously done for the MB moments is now equivalently carried out for the new $\mathcal{M}$ moment system. 
Hence, the advantages of the \textit{entire} new proposed construction are now outlined in details,  
in order to answer the question in the second issue at the end of section \ref{section:IntroductionHigherOrder}. 

Similarly to the MB coefficients, the $\mathcal{M}$ coefficients are also denoted here as $Q_i$, $R_i$, $S_i$ and $V_i$, which correspond to each $u^i$ velocity terms in Eqs. \eqref{eq:AccuracyD1Q51D-3}-\eqref{eq:AccuracyD1Q51D-5} and \eqref{eq:AccuracyD1Q71D-6} respectively. The term $\theta = R T$, found in the MB moment Eq. \eqref{eq:AccuracyD1Q51D-2} and in the lattice kinematic viscosity, becomes $R T (1 + 2 \mu)$ in  the $\mathcal{M}$ moment system (c.f. section \ref{section:GeneralizedHermite}). That is, a sort of a rescaled $T$ value, seen from an algebraic point of view. The rest of the $\theta$-linked coefficients, e.g. $Q_1$, $R_0$, $R_2$, $R_2$, $S_1$, $S_3$, $V_0$, $V_2$, $V_4$ c.f. Eqs. \eqref{eq:AccuracyD1Q51D-3}-\eqref{eq:AccuracyD1Q51D-5} and \eqref{eq:AccuracyD1Q71D-6}, are subjected to equivalent transformations, c.f. table \ref{tab:GeneralizedCoefficientsD1QnqMU}. 
A comparison between tables \ref{tab:MBCoefficientsD1Qnq} and \ref{tab:GeneralizedCoefficientsD1QnqMU} reveals that the $\theta$-linked MB coefficients $a \cdot \theta^n$ become $b \cdot \theta^n \cdot (c + 2\mu) \cdot (d + 2\mu) \cdot \dots $, $n$-times in the corresponding $\mathcal{M}$ coefficients, where $a = b \cdot c \cdot d \cdot \dots$ and $c$, $d$, etc are  positive odd integers. 
The other coefficients, e.g.  $Q_3$, $R_4$, $S_5$ and $V_6$, are the same as the MB coefficients. 

The obtained new $\mathcal{M}$ moments, whose some coefficients are found in table \ref{tab:GeneralizedCoefficientsD1QnqMU}, can be directly generated in a similar way as the MB moments are acquired from Eq. \eqref{eq:MB-convective-moments}. A way can be to use the following relation
\begin{eqnarray} \label{eq:Machado-convective-moments}
\sum_{i=0}^{n_q-1} f_i^{\textrm{eq}} c^{(M)} =   \rho \ \textrm{\Large{\textit{e}}}\Big(-\frac{u^2}{2 \mathcal{F}_M \theta}\Big) (\mathcal{F}_M \ \theta)^M 
\frac{\partial^M }{\partial u^M} \Bigg(  \textrm{\Large{\textit{e}}}\Big(\frac{u^2}{2 \mathcal{F}_M  \theta}\Big)  \Bigg),  
\end{eqnarray}
from which the generated moments (rhs in \eqref{eq:Machado-convective-moments}) are expanded and the containing terms $\mathcal{F}_a^b$  are subsequently substituted by  
\begin{eqnarray} \label{eq:FunctionWithMod}
\mathcal{F}_{n}^{m_{\textrm{Max}}} = \prod_{m=0}^{m_{\textrm{Max}}-1} \frac{n-2m-\textrm{mod}(n-2m+1,2)+2 \mu}{n-2m-\textrm{mod}(n-2m+1,2)}.
\end{eqnarray}
The term $\textrm{mod}(i,j)=k$ represents the modulo operation, where $k$ is the reminder on division $i/j$.  
Note from Eq. \eqref{eq:FunctionWithMod} that $\mathcal{F}_{n}^{m_{\textrm{Max}}}=1$ when $\mu=0$ and then Eq.  \eqref{eq:Machado-convective-moments} gives the MB moments (c.f. Eq. \eqref{eq:MB-convective-moments}). 

\begin{figure}[!htbp]
 \begin{center}
 { \scriptsize
\includegraphics[angle=0, width=1.15\textwidth]{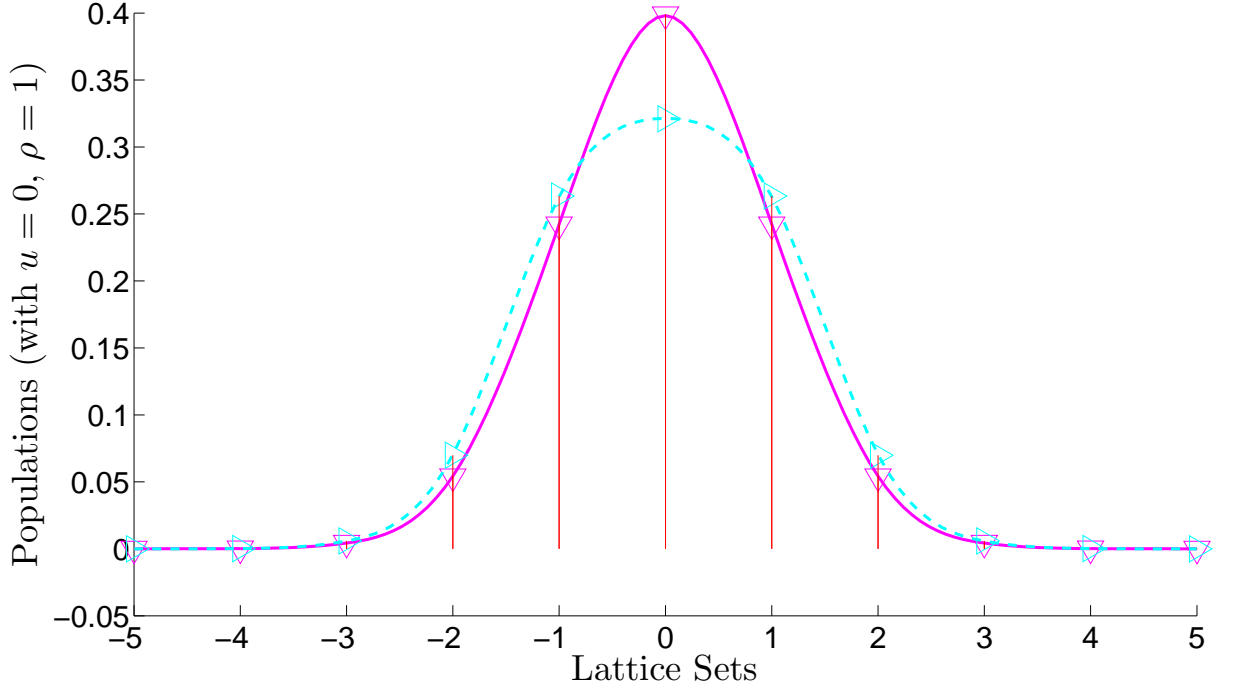}
    \caption{(Color online) Weights values (i.e. populations  $f_i^{\textrm{eq}}$ with $u=0$ and $\rho=1$) and the \textit{likely} shapes of their distributions for the D1Q11 with $\theta=1.0$. \textcolor{magenta}{\Large{$\triangledown$} $-$} (triangle down-solid): $\mu=0$; \textcolor{cyan}{\Large{$\triangleright$} - -} (triangle right-dashed): $\mu=1/10$.
    }
    \label{fig:Weights1DmuGeneralizedHigherOrderD1Q11MuZeroMu1Div10}
    }
  \end{center}
\end{figure}

The equivalent procedure used to determine Eq. \eqref{eq:ReferenceTemperatureD1Qnq} is carried out now to obtain $\theta = \theta_0$ from 
\begin{eqnarray} \label{eq:GeneralizedReferenceTemperatureD1Qnq} 
\sum_{k=0}^z (-1)^{k} \ 2^{z+i-k} (\mu + \frac{1}{2})_{z+1-k} \ \theta^{z-k} e_k(c_1^2, c_2^2, \dots, c_z^2) = 0, 
\end{eqnarray} 
for the D1Q3, D1Q5, D1Q7 and D1Q9 models, while for the D1Q11 model, the term 
$- 36 (5 + 2 \mu) (3 + 2 \mu)$ is added into the $\theta^5$ part of the polynomial generated by \eqref{eq:GeneralizedReferenceTemperatureD1Qnq}, from which the roots are obtained.  
The terms $(\mu + \frac{1}{2})_{z+1-k}$ and $ e_k(c_1^2, c_2^2, \dots, c_z^2) $ are the Pochhammer symbol and the $k$th-elementary symmetric polynomial respectively. 
When $\mu=0$, the term $2^{z+i-k} (\mu + \frac{1}{2})_{z+1-k}$ and Eq. \eqref{eq:GeneralizedReferenceTemperatureD1Qnq} become $(n_q-2\cdot k)!!$ and Eq. \eqref{eq:ReferenceTemperatureD1Qnq} respectively. 
The final results are presented in table \ref{tab:GeneralizedCoefficientsD1QnqMU} and some examples in table \ref{tabular:GeneralizedReferenceTemperatureD1Qnq}, which become equal to those in tables \ref{tab:MBCoefficientsD1Qnq} and  \ref{tabular:ReferenceTemperatureD1Qnq} respectively when $\mu=0$, and has been acquired following the same procedure around the Eqs. \eqref{eq:MBcoefficientForD1Q5-Q3}-\eqref{eq:ReferenceTemperatureD1Q5}. 
The example values of $\theta_0$ in table \ref{tabular:GeneralizedReferenceTemperatureD1Qnq} are presented in that way (around machine precision) so they can be compared to their corresponding $\theta_0$ values in table \ref{tabular:ReferenceTemperatureD1Qnq}.  Note that the $\theta_0$ values in table \ref{tabular:GeneralizedReferenceTemperatureD1Qnq} are lower than those in table \ref{tabular:ReferenceTemperatureD1Qnq}.  The peakedness of the distribution is affected when $\mu=0$ is increased to some value $\mu \neq 0$, c.f. Fig. \ref {fig:Weights1DmuGeneralizedHigherOrderD1Q11MuZeroMu1Div10}, where the distribution of the D1Q11 model is depicted with $\theta=1.0$, $\rho=1.0$ and $u=0$, for both $\mu=0$ and $\mu=1/10$. 
Long tailed and subexponentials distributions can also be obtained for high $z$ values when $\mu \neq 0$.

\begin{table}[!htbp]
	\centering
	{ \footnotesize
\begin{tabular}{|p{0.95\linewidth}|} 
\hline 
			\begin{eqnarray} \label{eq:RefertenceTemperatureD1Q5WithC2-2MU} 
			\theta_0( \textrm{ with } \boldsymbol{c}=\{0, \pm 1, \pm 2	 \} \textrm{ with } \mu=1/3) &=& \frac{15}{34}+\frac{3}{374} \sqrt{33}.
			\end{eqnarray}
\\
\hline 	 
			\begin{eqnarray} \label{eq:RefertenceTemperatureD1Q7WithC3-3MU}
			\theta_0( \textrm{ with } \boldsymbol{c}=\{0, \pm 1, \pm 2	\pm 3	 \} \textrm{ and } \mu=1/5) &=& 
			0.498 \ 011 \ 143 \ 151 \ 771 \ 857 \ 6.
			\end{eqnarray}
\\
\hline 
\begin{eqnarray} \label{eq:RefertenceTemperatureD1Q9WithC4MU}
	\theta_0 (\textrm{ with } \boldsymbol{c}=\{0, \pm 1, \pm 2, \pm 3, \pm 4	 \} \textrm{ and } \mu=1/5) &=& 
	0.531 \ 822 \ 832 \ 492 \ 398 \ 970 \ 86.
 \end{eqnarray}
\\
\hline  
\begin{eqnarray} \label{eq:RefertenceTemperatureD1Q11WithC5-5MU}
	\theta_0 (\textrm{ with } \boldsymbol{c}=\{0, \pm 1, \pm 2, \pm 3, \pm 4, \pm 5	 \} \textrm{ and } \mu=1/10) &=& 
	2.056 \ 245 \ 985 \ 122 \ 330 \ 338 \ 8. \qquad
\end{eqnarray}
\\
\hline 
\end{tabular}
	}
\caption{Examples of reference values of $\theta_0 = $ function$(c_i, \mu)$ for some one-dimensional  D1Q$n_q$ models (c.f. Eq. \eqref{eq:GeneralizedReferenceTemperatureD1Qnq}) and shortest \textit{on}-Cartesian lattice velocity sets $\boldsymbol{c} = \{0, \pm c_1, \dots, \pm c_z \}$, so the $\mathcal{M}$ $(z+1)$-moment is completely fulfilled.
Eq. \eqref{eq:RefertenceTemperatureD1Q5WithC2-2MU}: D1Q5; Eq. \eqref{eq:RefertenceTemperatureD1Q7WithC3-3MU}: D1Q7; Eq. \eqref{eq:RefertenceTemperatureD1Q9WithC4MU}: D1Q9; Eqs. \eqref{eq:RefertenceTemperatureD1Q11WithC5-5MU}: D1Q11. All populations \eqref{eq1:HermiteMu-EDF} with $\rho=1.0$ and weights \eqref{eq:1DmuGeneralizedWeights} are positive with given $\mu$ and $\theta = \theta_0$ values provided that: D1Q5: $0 \leq |u| \leq 0.802$; D1Q7: $0 \leq |u| \leq 1.081$; D1Q9: $0 \leq |u| \leq 0.443$; D1Q11: $0 \leq |u| \leq 1.323$.}\label{tabular:GeneralizedReferenceTemperatureD1Qnq}
\end{table}

The results obtained from the shortest lattice (the most desirable lattice due to their ``more local'' property) are presented in table \ref{tabular:GeneralizedReferenceTemperatureD1Qnq} with $\mu \neq 0$.  Unlike with the MB moments, the $\mathcal{M}$ moment system gives $\theta_0 = $ function$(c_i, \mu)$, c.f. Eqs. \eqref{eq:ReferenceTemperatureD1Qnq}  and \eqref{eq:GeneralizedReferenceTemperatureD1Qnq}, where the extra $\mu$ parameter can give the theoretical possibility to obtain the shortest lattice with a real-valued $\theta_0$. Hence, the lattice should not longer solely blamed for the existence of complex $\theta_0$ values when $\theta_0 = $ function$(c_i, \mu)$. For example, with lattice velocity integers $\{ 0, \pm 1, \pm 2 \}$ and $\{0, \pm 1, \pm 2, \pm 3 , \pm 4  \}$, the complex valued $\theta_0$ in Eqs. \eqref{eq:RefertenceTemperatureD1Q5WithC2-2} and \eqref{eq:RefertenceTemperatureD1Q9WithC4} become real valued in Eqs. \eqref{eq:RefertenceTemperatureD1Q5WithC2-2MU} and \eqref{eq:RefertenceTemperatureD1Q9WithC4MU} respectively with $\mu \neq 0$.  All this while the previous advantages acquired from the use of the MB moments are kept, i.e. the first $\mathcal{M}$ $z$-moments are thermally matched with the use of the $\mu$-generalized Hermite-based LB construction \textit{on}-Cartesian lattice. The $\mathcal{M}$ $(z+1)$-moment is isothermally fulfilled with the shortest \textit{on}-Cartesian lattice set. This is clearly an advantage of the new proposed LB construction, compared to the previous Hermite and ELB models.

It has already mentioned in this work that the theoretical valid range of example of $\theta$ values, useful in the first hydrodynamic $z$-moments, can be obtained from Eqs. \eqref{eq:MaxTemp} so that the thermal weights (in the EDF) are non-negative. However, only the valid range of $\theta$ for the low-order D1Q3 model is given so far (c.f. section  \ref{section:GeneralizedHermite}). LB equations are discrete formulations, and thus the possible values of $\theta$ can be segmented. Only the largest ranges for each case are presented. The largest valid ranges of $\theta$ for the D1Q5, D1Q7, D1Q9 and D1Q11 lattice models are given in table \ref{tabular:ValidRangeMuZeroAndGeneralizedTemperatureD1Qnq}. It should be noted that the reference $\theta_0$ values in tables \ref{tabular:ReferenceTemperatureD1Qnq} and \ref{tabular:GeneralizedReferenceTemperatureD1Qnq} are within the theoretical valid ranges of $\theta$ given in table \ref{tabular:ValidRangeMuZeroAndGeneralizedTemperatureD1Qnq}. Based on tables \ref{tabular:ReferenceTemperatureD1Qnq}  and  \ref{tabular:ValidRangeMuZeroAndGeneralizedTemperatureD1Qnq}, it is interesting to note that for $\mu=0$, the $\theta_0$ value of the D1Q(5+$n$) lattice can become one of a extreme $\theta_{\textrm{min/max}}$ value  for the next D1Q(5+$n$+2) lattice model, where $n=0, 2$ and $4$, c.f. Eqs. $\theta_0=1/3$ (in section \ref{section:GeneralizedHermite}),  \eqref{eq:RefertenceTemperatureD1Q5WithC2-3}, \eqref{eq:RefertenceTemperatureD1Q7WithC3-3}, \eqref{eq:RefertenceTemperatureD1Q9WithC5} and \eqref{eq:ValidTemperatureD1Q5WithMuZeroMin}, \eqref{eq:ValidTemperatureD1Q7WithMuZeroMax}, \eqref{eq:ValidTemperatureD1Q9WithMuZeroMin}, \eqref{eq:ValidTemperatureD1Q11WithMuZeroMin}   respectively. Similar findings can be observed for some $\mu \neq 0$ cases, c.f. Eqs. \eqref{eq:RefertenceTemperatureD1Q7WithC3-3MU} and \eqref{eq:ValidTemperatureD1Q9WithMuMin}.


\begin{table}[!htbp]
	\centering
	{ \footnotesize
\begin{tabular}{|p{0.95\linewidth}|} 
\hline 
\begin{subequations} \label{eq:ValidTemperatureD1Q5WithMuZero}
			\begin{eqnarray} \label{eq:ValidTemperatureD1Q5WithMuZeroMin} 
			\theta_{\textrm{min}}( \textrm{ with } \boldsymbol{c}=\{0, \pm 1, \pm 3\} \textrm{ and } \mu=0)  &=& \frac{1}{3}. \\ \label{eq:ValidTemperatureD1Q5WithMuZeroMax}
			\theta_{\textrm{max}}( \textrm{ with } \boldsymbol{c}=\{0, \pm 1, \pm 3\} \textrm{ and } \mu=0)  &=& 3.
			\end{eqnarray}
\end{subequations} 
\\
\hline 	 
\begin{subequations} \label{eq:ValidTemperatureD1Q5WithMu}
			\begin{eqnarray} \label{eq:ValidTemperatureD1Q5WithMuMin} 
			\theta_{\textrm{min}}( \textrm{ with } \boldsymbol{c}=\{0, \pm 1, \pm 2\} \textrm{ and } \mu=1/3)  &=& \frac{3}{11}. \\ \label{eq:ValidTemperatureD1Q5WithMuMax}
			\theta_{\textrm{max}}( \textrm{ with } \boldsymbol{c}=\{0, \pm 1, \pm 2\} \textrm{ and } \mu=1/3)  &=& \frac{12}{11}.
			\end{eqnarray}
\end{subequations} 
\\
\hline 	 
\begin{subequations} \label{eq:ValidTemperatureD1Q7WithMuZero}
			\begin{eqnarray} \label{eq:ValidTemperatureD1Q7WithMuZeroMin} 
			\theta_{\textrm{min}}( \textrm{ with } \boldsymbol{c}=\{0, \pm 1, \pm 2, \pm 3\} \textrm{ and } \mu=0)  &=& 1-\frac{\sqrt{10}}{5}. \\ \label{eq:ValidTemperatureD1Q7WithMuZeroMax}
			\theta_{\textrm{max}}( \textrm{ with } \boldsymbol{c}=\{0, \pm 1, \pm 2, \pm 3\} \textrm{ and } \mu=0)  &=& 1+\frac{\sqrt{10}}{5}.
			\end{eqnarray}
\end{subequations} 
\\
\hline 	 
\begin{subequations} \label{eq:ValidTemperatureD1Q7WithMu}
			\begin{eqnarray} \label{eq:ValidTemperatureD1Q7WithMuMin} 
			\theta_{\textrm{min}}( \textrm{ with } \boldsymbol{c}=\{0, \pm 1, \pm 2, \pm 3\} \textrm{ and } \mu=1/5)  &=& \frac{25}{27} -5 \frac{\sqrt{3094}}{459}. \\ \label{eq:ValidTemperatureD1Q7WithMuMax}
			\theta_{\textrm{max}}( \textrm{ with } \boldsymbol{c}=\{0, \pm 1, \pm 2, \pm 3\} \textrm{ and } \mu=1/5)  &=& 1.401 \ 283 \ 831 \ 980 \ 340 \ 563 \ 9.
			\end{eqnarray}
\end{subequations} 
\\
\hline 	 
\begin{subequations} \label{eq:ValidTemperatureD1Q9WithMuZero}
			\begin{eqnarray} \label{eq:ValidTemperatureD1Q9WithMuZeroMin} 
			\theta_{\textrm{min}}( \textrm{ with } \boldsymbol{c}=\{0, \pm 1, \pm 2, \pm 3, \pm 5\} \textrm{ and } \mu=0)  &=& 0.697 \ 953 \ 322 \ 019 \ 683 \ 088 \ 24. \qquad \\ \label{eq:ValidTemperatureD1Q9WithMuZeroMax}
			\theta_{\textrm{max}}( \textrm{ with } \boldsymbol{c}=\{0, \pm 1, \pm 2, \pm 3, \pm 5\} \textrm{ and } \mu=0)  &=& 2.881 \ 311 \ 061 \ 716 \ 039 \ 428 \ 2. \qquad
			\end{eqnarray}
\end{subequations} 
\\
\hline 	 
\begin{subequations} \label{eq:ValidTemperatureD1Q9WithMu}
			\begin{eqnarray} \label{eq:ValidTemperatureD1Q9WithMuMin} 
			\theta_{\textrm{min}}( \textrm{ with } \boldsymbol{c}=\{0, \pm 1, \pm 2, \pm 3, \pm 4 \} \textrm{ and } \mu=1/5)  &=& 0.498 \ 011 \ 143 \ 151 \ 771 \ 857 \ 6. \qquad \\ \label{eq:ValidTemperatureD1Q9WithMuMax}
			\theta_{\textrm{max}}( \textrm{ with } \boldsymbol{c}=\{0, \pm 1, \pm 2, \pm 3, \pm 4\} \textrm{ and } \mu=1/5)  &=& 1.829 \ 636 \ 973 \ 881 \ 101 \ 141 \ 2. \qquad
			\end{eqnarray}
\end{subequations} 
\\
\hline 	 
\begin{subequations} \label{eq:ValidTemperatureD1Q11WithMuZero}
			\begin{eqnarray} \label{eq:ValidTemperatureD1Q11WithMuZeroMin} 
			\theta_{\textrm{min}}( \textrm{ with } \boldsymbol{c}=\{0, \pm 1, \pm 2, \pm 3, \pm 4, \pm 5\} \textrm{ and } \mu=0)  &=& 0.756 \ 080 \ 852 \ 594 \ 268 \ 582 \ 31. \qquad \\ \label{eq:ValidTemperatureD1Q11WithMuZeroMax}
			\theta_{\textrm{max}}( \textrm{ with } \boldsymbol{c}=\{0, \pm 1, \pm 2, \pm 3, \pm 4, \pm 5\} \textrm{ and } \mu=0)  &=& 2.175 \ 382 \ 386 \ 573 \ 040 \ 694 \ 7. \qquad
			\end{eqnarray}
\end{subequations} 
\\
\hline 	 
\begin{subequations} \label{eq:ValidTemperatureD1Q11WithMu}
			\begin{eqnarray} \label{eq:ValidTemperatureD1Q11WithMuMin} 
			\theta_{\textrm{min}}( \textrm{ with } \boldsymbol{c}=\{0, \pm 1, \pm 2, \pm 3, \pm 4, \pm 5 \} \textrm{ and } \mu=1/10)  &=& 0.963 \ 908 \ 781 \ 629 \ 469 \ 643. \qquad \\ \label{eq:ValidTemperatureD1Q11WithMuMax}
			\theta_{\textrm{max}}( \textrm{ with } \boldsymbol{c}=\{0, \pm 1, \pm 2, \pm 3, \pm 4, \pm 5\} \textrm{ and } \mu=1/10)  &=& 2.141 \ 493 \ 081 \ 363 \ 463 \ 722. \qquad
			\end{eqnarray}
\end{subequations} 
\\
\hline 	 
\end{tabular}
	}
\caption{Theoretical largest valid range of examples of $\theta$ for the D1Q5 (Eqs. \eqref{eq:ValidTemperatureD1Q5WithMuZero}-\eqref{eq:ValidTemperatureD1Q5WithMu}), D1Q7 (Eqs. \eqref{eq:ValidTemperatureD1Q7WithMuZero}-\eqref{eq:ValidTemperatureD1Q7WithMu}), D1Q9 (Eqs. \eqref{eq:ValidTemperatureD1Q9WithMuZero}-\eqref{eq:ValidTemperatureD1Q9WithMu}) and D1Q11 (Eqs. \eqref{eq:ValidTemperatureD1Q11WithMuZero}-\eqref{eq:ValidTemperatureD1Q11WithMu}) lattice models so that the weights are positive, c.f. Eqs. \eqref{eq:MaxTemp}. The extremes should be excluded, i.e. $\theta = ]\theta_{\textrm{min}},\theta_{\textrm{max}} [$.}\label{tabular:ValidRangeMuZeroAndGeneralizedTemperatureD1Qnq}
\end{table}

Eq. \eqref{eq:IsothermalNavier-Stokes} is presented for $\theta =$ constant, but the fixed $\theta$ value is not specified. 
It can be argued that $\theta_1=(1+2 \mu) \theta_2$ algebraically, for non trivial values. 
Although there exist many $\theta_1$, $\mu$ and $\theta_2$  values, some few concrete examples are now outlined. 
For the D1Q7 case: With 
$\theta_1=0.69721560041248060075$, 
which is within the extremes \eqref{eq:ValidTemperatureD1Q7WithMuZero} in table \ref{tabular:ValidRangeMuZeroAndGeneralizedTemperatureD1Qnq}, 
and $\mu=1/5$, leads to a $\theta_2$ equal to the reference value \eqref{eq:RefertenceTemperatureD1Q7WithC3-3MU} in table \ref{tabular:GeneralizedReferenceTemperatureD1Qnq}. For the D1Q9 case: With 
$\theta_1=0.7445519654893585592$, 
which is within the extremes \eqref{eq:ValidTemperatureD1Q9WithMuZero} in table \ref{tabular:ValidRangeMuZeroAndGeneralizedTemperatureD1Qnq}, 
and $\mu=1/5$, leads to a $\theta_2$ equal to the reference value \eqref{eq:RefertenceTemperatureD1Q9WithC4MU} in table \ref{tabular:GeneralizedReferenceTemperatureD1Qnq}. For the D1Q11 case: $\theta_1=1.8$, which is within the extremes \eqref{eq:ValidTemperatureD1Q11WithMuZero} in table \ref{tabular:ValidRangeMuZeroAndGeneralizedTemperatureD1Qnq}, and $\mu=1/10$, leads to $\theta_2=1.5$, which is within the extremes \eqref{eq:ValidTemperatureD1Q11WithMu} in table \ref{tabular:ValidRangeMuZeroAndGeneralizedTemperatureD1Qnq}. 
Hence, exactly same results can be theoretically obtained from the Eq. \eqref{eq:IsothermalNavier-Stokes}, for both MB and $\mathcal{M}$ moment systems. 
The D1Q5 case can be excluded because it only 
matches the fourth ($M=3$) hydrodynamic moment (i.e. complete Galilean invariant) 
at $\theta=\theta_0$. However,  
with a chosen appropriate flow velocity so that $u^3 \approx 0$ it is observed that with $\theta_1=25/34+5 \sqrt{33}/374$, which is within the extremes \eqref{eq:ValidTemperatureD1Q5WithMuZero} in table \ref{tabular:ValidRangeMuZeroAndGeneralizedTemperatureD1Qnq}, and $\mu=1/3$, leads to a $\theta_2$ equal to the reference value \eqref{eq:RefertenceTemperatureD1Q5WithC2-2MU} in table \ref{tabular:GeneralizedReferenceTemperatureD1Qnq}. 
This $\theta_2=\theta_0$ value is needed 
to match the fourth hydrodynamic $\mathcal{M}$ moment,   
with the shortest \textit{on}-Cartesian lattice set. 

\begin{figure}
\centering
	\begin{center}
	 { \scriptsize
\begin{tabular}{cc}
\epsfig{file=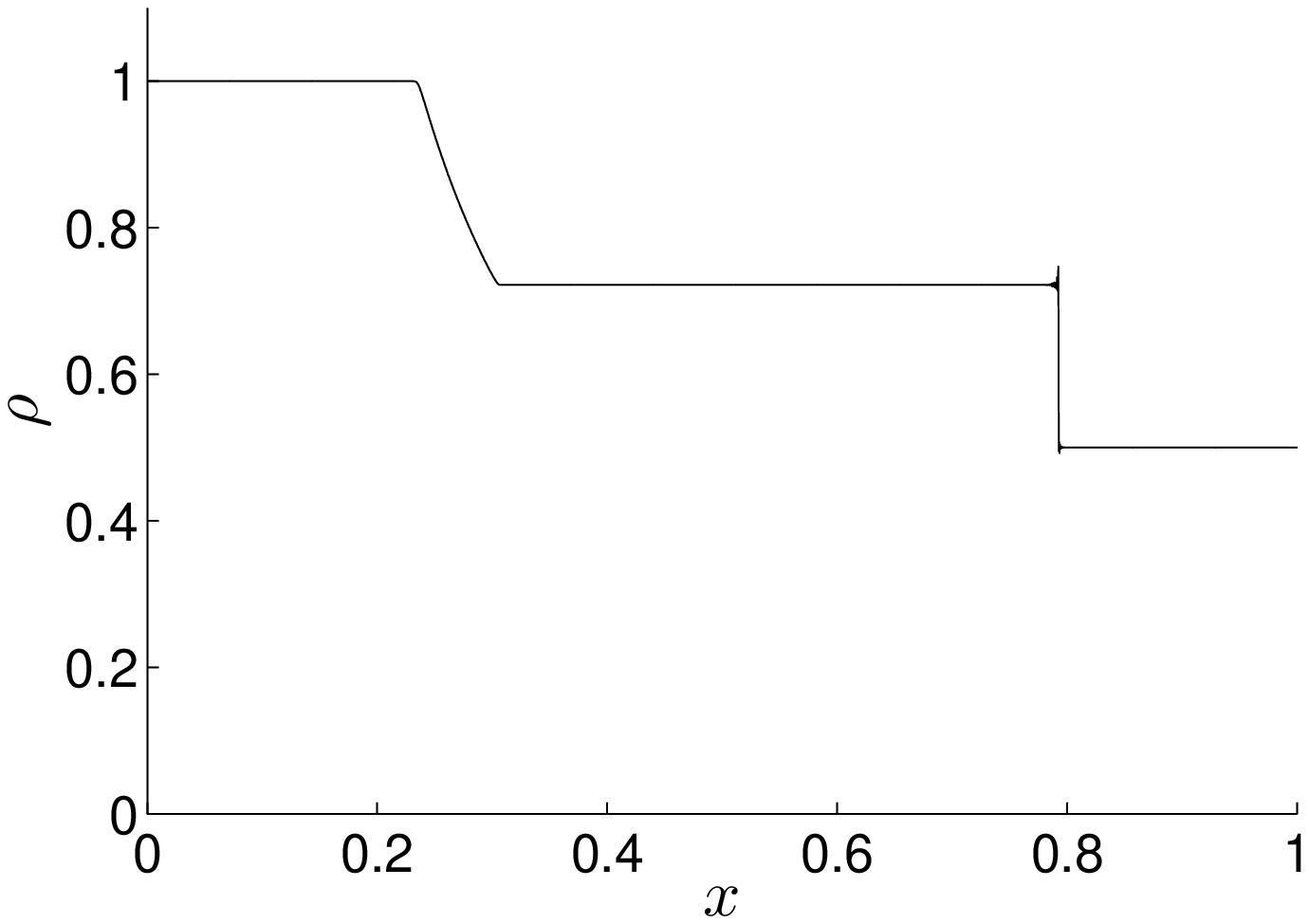,angle=0,width=0.5\linewidth,clip=}  &  \epsfig{file=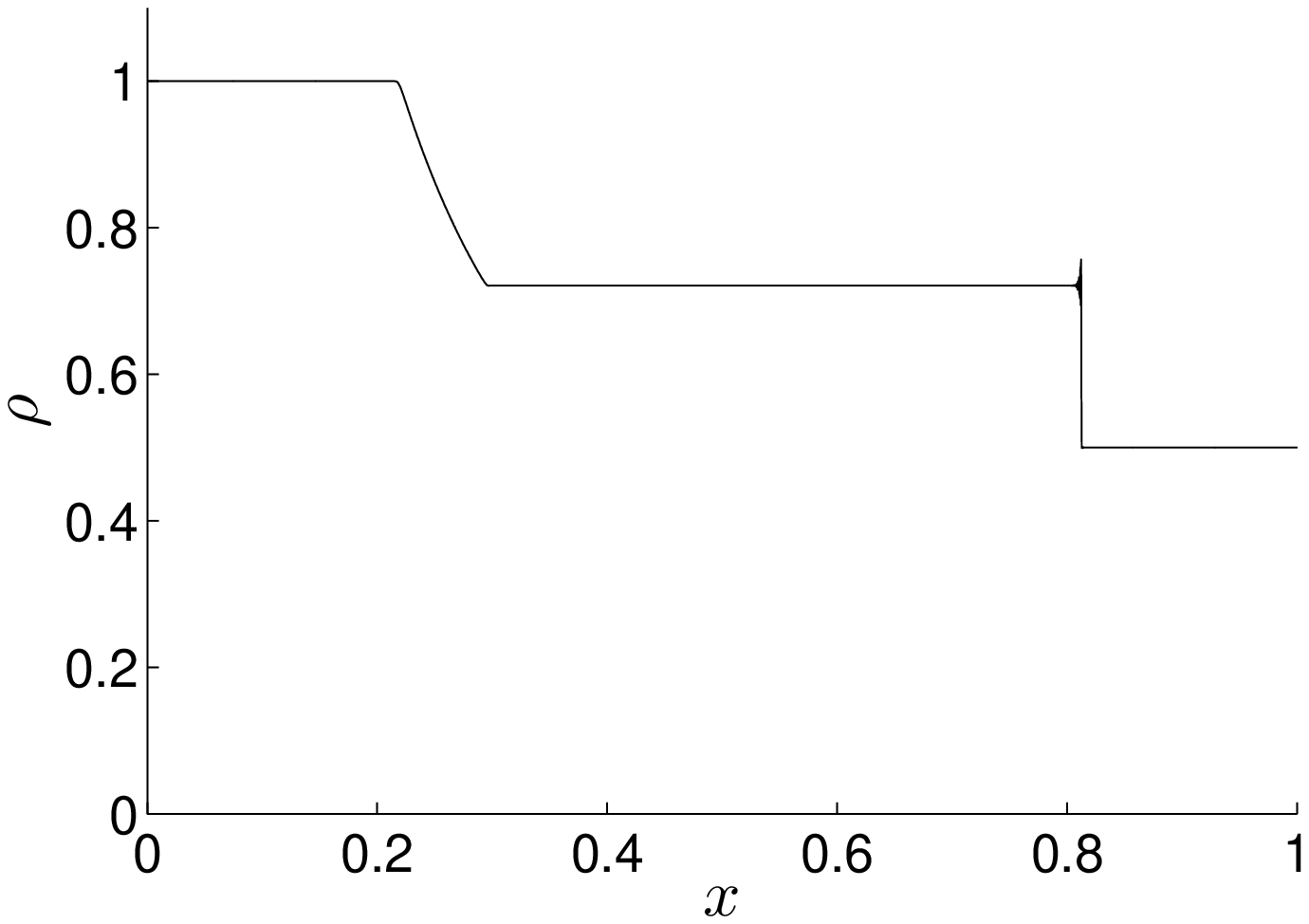,angle=0,width=0.5\linewidth,clip=} \\ 
a)  & b)  \\
\end{tabular}
    \caption{(Color online) Simulation of the one-dimensional shock tube problem by the $\mu$-generalized Hermite high-order LB construction for two D1Q9 cases with kinematic viscosity $\nu=1/30$: a) $\mu=0$, lattice set $\{0, \pm 1, \pm 2, \pm 3, \pm 5 \}$ and $\theta_0$ found in \eqref{eq:RefertenceTemperatureD1Q9WithC5}, table \ref{tabular:ReferenceTemperatureD1Qnq}; b) $\mu=1/5$, shortest \textit{on}-Cartesian lattice set $\{0, \pm 1, \pm 2, \pm 3, \pm 4 \}$ and real-valued $\theta_0$ found in \eqref{eq:RefertenceTemperatureD1Q9WithC4MU}, table  \ref{tabular:GeneralizedReferenceTemperatureD1Qnq}. Length of the domain $8 \times 10^p$, time step $3 \times 10^p$, $p=3$. 
    }
    \label{fig:1DShockTubeD1Q9MuZeroMu1div5}
    }
	\end{center}
\end{figure}

Although this work is predominantly theoretical a numerical test is presented, to demonstrate the feasibility of the proposed $\mu$-generalized Hermite high-order LB construction for both $\mu=0$ and $\mu \neq 0$. Two D1Q9 cases are chosen: a) with  lattice set $\{0, \pm 1, \pm 2, \pm 3, \pm 5 \}$, $\mu=0$, reference value $\theta_0$ found in \eqref{eq:RefertenceTemperatureD1Q9WithC5}, table \ref{tabular:ReferenceTemperatureD1Qnq}; b) with shortest lattice set $\{0, \pm 1, \pm 2, \pm 3, \pm 4 \}$, $\mu=1/5$ and reference value $\theta_0$ found in \eqref{eq:RefertenceTemperatureD1Q9WithC4MU}, table \ref{tabular:GeneralizedReferenceTemperatureD1Qnq}. With kinematic viscosity $\nu = (\tau - 1/2) (1+2 \mu) \theta_0 = 1/30$ and their corresponding values of $\mu$ and $\theta_0$, the $\tau$ values needed in the LBGK formulation for each case are obtained. A one-dimensional shock tube is simulated with an initial density ratio of 1:2 so that $\rho=1.0$ for $x \leq L/2$, $L$ being the length of the domain, and $\rho=0.5$ otherwise. 
The results are depicted  at the same time step in Fig. \ref{fig:1DShockTubeD1Q9MuZeroMu1div5}. 
In Fig. \ref{fig:1DShockTubeD1Q9MuZeroMu1div5}, a compressible front moving into the low-density region while a rarefaction front moving into a high-density region are observed, as expected from these kinds of simulations. 
The observed oscillatory pattern at the shock is common in the lattice Boltzmann schemes, c.f.  \cite{BrownleeGorbanLevesley2008}, \cite{ChikatamarlaKarlin1D2006}.  
Both D1Q9 cases are \textit{on}-Cartesian LB models, but the one with $\mu \neq 0$, Fig. \ref{fig:1DShockTubeD1Q9MuZeroMu1div5}  b), has the shortest lattice set $\{0, \pm 1, \pm 2, \pm 3, \pm 4 \}$. The sole purpose of this numerical test is for a simple  computational proof of concept. Further numerical studies are carried out elsewhere.

Certainly, it would be interesting to obtain more hydrodynamic terms from the $\mathcal{M}$ moment system beside $p=\rho R T (1+2 \mu)$ and $\nu=(\tau-1/2) R T (1+2 \mu)$.  However, $i$) it would derail this 
 work from its two main general issues, mentioned in the abstract and in the last paragraphs at the end of section \ref{section:IntroductionHigherOrder}, $ii$) 
 such study has to be placed within the context of another set of references 
 because based on previous works (c.f. \cite{Humiere1992}, \cite{HeChenDoolenJCP1998}, \cite{ShanChen2007}, \cite{PhilippiHegeleSurmasSiebertSantos2007}, \cite{LSLuoICMMES2007}, \cite{NieShanChen2008}) another LB formulation would be needed to compensate the limitations of the LBGK. This deserves a separate work and it is presented by the author elsewhere. 

\begin{table}
	\centering
		{ \small 											
		\begin{tabular}[t]{|ccccccccccc|l}
		\hline
		 High-order LB  & \vline & $i$) \textit{on}-Cartesian  & \vline & $ii$) Thermal & \vline & $iii$) Thermal & \vline &  $iv$) Spurious  & \vline & $v$) Shortest lattice \\ 
		 constructions         & \vline & lattice  & \vline &  moments & \vline & weights & \vline &  velocity terms & \vline &  sets \\ 
		\hline
		 ELB construction, c.f. \cite{ChikatamarlaKarlin2009},  & \vline & \textcolor{green}{\underline{\textcolor{black}{Yes}}} & \vline & No & \vline & No & \vline &  Yes & \vline & No \\
		 Refs. therein and appendix.  & \vline & & \vline &  & \vline & & \vline &   & \vline &  \\
		\hline
		Previous Hermite    & \vline & & \vline &  & \vline & & \vline &   & \vline &  \\
		construction, c.f. \cite{Shan2010} & \vline & \textcolor{green}{\underline{\textcolor{black}{Yes}}} & \vline & \textcolor{green}{\underline{\textcolor{black}{Yes}}} & \vline & No & \vline &       \textcolor{green}{\underline{\textcolor{black}{No}}} & \vline &  No \\
		 and references therein.  & \vline & & \vline &  & \vline & & \vline &  & \vline &  \\ 
		\hline
\multicolumn{1}{|c}{\multirow{2}{*}{}} & 
\multicolumn{1}{|c|}{}  &  & \vline &  & \vline &  & \vline &  & \vline &  \\ 
\multicolumn{1}{|c}{\multirow{2}{*}{Proposed}} &
\multicolumn{1}{|c|}{  with $\mu = 0$  }  & \textcolor{green}{\underline{\textcolor{black}{Yes}}} & \vline & \textcolor{green}{\underline{\textcolor{black}{Yes}}} & \vline & \textcolor{green}{\underline{\textcolor{black}{Yes}}} & \vline & \textcolor{green}{\underline{\textcolor{black}{No}}} & \vline & No \\ 
\multicolumn{1}{|c}{\multirow{2}{*}{Hermite}} &
\multicolumn{1}{|c|}{}  &  & \vline &  & \vline &  & \vline &  & \vline &  \\  \cline{2-11}
\multicolumn{1}{|c}{\multirow{2}{*}{construction}} &
\multicolumn{1}{|c|}{}  &  & \vline &  & \vline &  & \vline &  & \vline &  \\ 
\multicolumn{1}{|c}{\multirow{2}{*}{}} & 
\multicolumn{1}{|c|}{ with $\mu \neq 0$ } & \textcolor{green}{\underline{\textcolor{black}{Yes}}} & \vline & \textcolor{green}{\underline{\textcolor{black}{Yes}}} & \vline & \textcolor{green}{\underline{\textcolor{black}{Yes}}} & \vline & \textcolor{green}{\underline{\textcolor{black}{No}}}  & \vline & \textcolor{green}{\underline{\textcolor{black}{Yes}}} \\ 
\multicolumn{1}{|c}{\multirow{2}{*}{}} &
\multicolumn{1}{|c|}{}  &  & \vline &  & \vline &  & \vline &  & \vline &  \\  
\hline
		\end{tabular}
		}
			\caption{ Comparison among non-mixed high-order LB constructions (theoretically free of finite difference scheme) based on whether or not they can have: $i$) lattice velocities \textit{on}-Cartesian, $ii$) their first hydrodynamic $z$-moments thermally fulfilled exactly, $iii$) thermal weights (based on the final results that are used in the EDF), $iv$) spurious velocity terms in their first hydrodynamic $z$-moments, $v$) a stable (one-dimensional) model with the shortest \textit{on}-Cartesian lattice sets (e.g. $c_i = $ consecutive integers, Fig. \ref{fig:Cartesian}) capable to exactly match the hydrodynamic $(z+1)$-moments with some fixed real-valued $\theta$. Positive properties are underlined.}	\label{tab:ComparisonLEBHermiteShanHermiteMachado} 
\end{table}

Summarizing, a comparison is made in table \ref{tab:ComparisonLEBHermiteShanHermiteMachado} among some high-order LB models. The positive properties are underlined. Obviously, the new proposed LB construction with $\mu \neq 0$ has the most (theoretical) advantages. It is noticed that the insertion of new advantageous properties are obtained whilst previous advantageous properties are kept.

\section{Conclusion}\label{section:ConclusionHigherOrder}

The $\mu$-generalized Hermite polynomials is proposed into the lattice Boltzmann (LB) approach, where $\mu \neq 1/2 - n$, $n=1/2, 1, 2, 3, \dots$. In the process, a new moment \textit{system} (denoted as $\mathcal{M}$) is proposed (c.f. Eqs. \ref{eq:Machado-convective-moments} and \ref{eq:FunctionWithMod}). The $\mathcal{M}$ moment system reduces to the Maxwell-Boltzmann (MB) moments when $\mu=0$. A new equilibrium distribution function (EDF) based on the $\mu$-generalized Hermite polynomials is also introduced (c.f. Eq. \eqref{eq1:HermiteMu-EDF}). The new proposed higher-order LB construction is constrained into the main LB idea (c.f. \cite{SucciBook}, \cite{SucciLecture2006},  \cite{SucciTurbulanceDNA2008}, \cite{BrownleeGorbanLevesley2008}). 
The new formulation is \textit{on}-Cartesian lattice sets, in order to avoid the (theoretical) need of interpolations, approximations or finite difference schemes. A single formulation for one-dimensional \textit{thermal} weights (based on the final results that are used in the EDF) is introduced in this work for an unlimited $n_q$ \textit{on}-Cartesian lattice grid points, where $n_q = 2 z +1$ and $z=1, 2, 3, \dots$ (c.f. Eqs. \eqref{eq:1DmuGeneralizedWeights} and Fig. \ref{fig:Cartesian}). 
This is in clear contrast to previous athermal weights (c.f. \cite{ShanYuanChen2006} (tables 1,2,3 therein), \cite{ShanChen2007} (table 1 therein), \cite{MengZhangJCP2011} (table 1 therein)).   
Two- and three-dimensional thermal weights can be obtained by mean of algebraic products of the one-dimensional thermal weights. The thermal term means in this work that the ``temperature'' $\theta$ does not have to be a fixed value, unlike in the isothermal case. A fixed ``temperature'' $\theta = \theta_0$ value is denoted as reference value. The EDF (c.f. Eq.  \eqref{eq2:HermiteMu-EDF}) is of the form $f_i^{\textrm{eq}}=\rho W_i(1+C)$, where $C$ is zero when the flow velocity ($u$) is zero, and the importance of the weights values $W_i$ is noticed. The flow velocity can be zero at the boundaries/walls, e.g. for a laminar channel flow.  
The possibility of having a high-order thermal LB model \textit{on}-Cartesian lattice with thermal weights ($W_i$), free of interpolations and finite different schemes, can be useful when dealing with boundaries, in particular for prospective models  treating ``heated'' walls. The first hydrodynamic $z$-moments, where $\sum_{i} f_i c_i^M$, $M= 0, 1, 2 \dots , z$, are exactly matched thermally when the aforementioned introduced formulation of the thermal weights within the proposed EDF is implemented. 

Another important issue to deal with is to obtain a high-order LB construction so that it is as local as possible, and thus efficient (parallel) computations can be carried out. In previous higher-order LB models, some one-dimensional lattice sets,  e.g. $n_q=5$, prove intrinsic unstable when the shortest \textit{on}-Cartesian lattice sets are used (c.f.  \cite{QianZhou1998}, \cite{Dellar2005}, \cite{ChikatamarlaKarlin1D2006}, \cite{ChikatamarlaKarlin2009}). In the proposed high-order LB construction with $\mu = 0$, some complex-valued $\theta$ are also obtained in those cases  (c.f. table \ref{tabular:ReferenceTemperatureD1Qnq}), as in \cite{ChikatamarlaKarlin2009}. However, since $\theta$ = function($\mu, c_i$) is obtained in the full scale proposed LB construction, this is changed when $\mu \neq 0$ (c.f. table  \ref{tabular:GeneralizedReferenceTemperatureD1Qnq} and Fig. \ref{fig:1DShockTubeD1Q9MuZeroMu1div5}). The new high-order LB formulation proposes a general construction to obtain real-valued $\theta$ using the \textit{shortest} \textit{on}-Cartesian lattice sets in one-dimension. Therefore, the highest hydrodynamic moments that can be exactly matched using the \textit{shortest} \textit{on}-Cartesian lattice sets in one-dimension in the proposed high-order LB construction, are the hydrodynamic $(z+1)$-moments, which are fulfilled isothermally (i.e. with $\theta = \theta_0$). 
Also, because of the thermal and accurate nature of the obtained relations, 
the presented approach is better than the one summarized in \cite{ChikatamarlaKarlin2009}, where a $z$-limited isothermal less accurate (with spurious velocity terms) construction is reported (c.f. a comparison in table  \ref{tab:ComparisonLEBHermiteShanHermiteMachado}). 
  
A single relation (c.f. Eqs. \eqref{eq:MaxTemp}), from which valid ranges of $\theta$ values can be extracted (c.f. table \ref{tabular:ValidRangeMuZeroAndGeneralizedTemperatureD1Qnq} for some ranges) so that the thermal weights are non-negative, is introduced.  
The reference $\theta_0$ values, needed to exactly match the hydrodynamic $(z+1)$-moments in the proposed LB construction (for both $\mu = 0$ and $\mu \neq 0$), for some of one-dimensional on-Cartesian lattice sets are provided (c.f. tables \ref{tabular:ReferenceTemperatureD1Qnq} and \ref{tabular:GeneralizedReferenceTemperatureD1Qnq}). These $\theta_0$ values can be obtained from a single relation (c.f. Eq. \eqref{eq:GeneralizedReferenceTemperatureD1Qnq}) for the D1Q$n_q$ models with $n_q=3-11$, which is also introduced in this work.  
Valid ranges of flow velocities ($u$) in lattice units for the D1Q$n_q$ models with $n_q=5-11$ with some fixed $\theta$ values, so that the populations are non-negative, are also presented (c.f. captions in tables \ref{tabular:ReferenceTemperatureD1Qnq},  \ref{tabular:GeneralizedReferenceTemperatureD1Qnq} and Fig. \ref{fig:PopulationsAndWeightsforD1Q11CasesHigherOrder}). 
It should be pointed out that proposed formulations for the thermal weights and the EFD (c.f. Eqs.  \eqref{eq:1DmuGeneralizedWeights} and \eqref{eq1:HermiteMu-EDF}) are presented in a general form, where the \textit{on}-Cartesian case is a particular one. 
A trade-off between the use of \textit{on}- and \textit{off}-Cartesian lattice sets when it comes to matching hydrodynamic coefficients is described by an example. 
The influence of the temperature, $z$ value and flow velocity on the \textit{likely} shapes of the distributions are also discussed. For high-order LB with very high $z$ values, the distributions can be long-tailed and subexponentials (c.f. Fig. \ref{fig:SubExponential}). Hence, the high-order LB construction is put on a firm theoretical ground (c.f.  \cite{EmbrechtsKluppelbergMikosch1997Book}), from which further theoretical studies can be carried out. 

The asked question in the abstract, introduction and section \ref{section:OnTheHigherOrder} is answered:  Yes, it is (theoretically) possible to obtain a high-order Hermite-based LB model able to fulfill the following three characteristics within a \textit{single} construction: 
capable to exactly match the first hydrodynamic $z$-moments thermally 
1) \textit{on}-Cartesian, 2) with \textit{thermal} weights (based on the final results that are used in the EDF), 3) 
whilst the hydrodynamic $(z+1)$-moments are exactly matched isothermally using the \textit{shortest} \textit{on}-Cartesian lattice sets.

\begin{acknowledgments}
Access to the University of Southampton software resources through VPN is sincerely acknowledged, which has enabled the realization of this work completely independently by the author from Sweden. Thanks to the managers of the Swedish Defence Research Agency (FOI) for their support. 
While waiting the reviewing process, this (limited version of the) paper is presented. 
\end{acknowledgments}



\section*{Appendix}

For completeness, self-consistent, \textit{some} of the main ideas behind construction of the ELB construction are summarized in this appendix. The derivation starts with a discrete $H$-function (Boltzmann ansatz) $H=(-S)=\sum_{i=0}^{n_q -1} f_i \textrm{log} (f_i/W_i )$, where $S$ is a concave function, which represents the entropy and $H$ is a convex function so that $H=-S$ \cite{WagnerAJ1998}, \cite{KarlinFerrante1999}, \cite{YongLiShiLuo2005}. The EDF is obtained by minimizing the $H$ function upon the constraints \eqref{eq:MB-convective-moments} up to $M=M_{\textrm{max}}$, where $M_{\textrm{max}}=1$ or to $M_{\textrm{max}}=2$, depending on the model, i.e.  
\begin{eqnarray} \label{eq:Minim-Equation} 
&&\frac{\partial H}{\partial f_i} + \sum_{M=0}^{M_{\textrm{max}}} \chi_{(M),\alpha}^* \frac{\partial ( \Phi(M)- \sum f_i c_{\alpha,i}^M) }{\partial f_i} 
=0,
\end{eqnarray}
where $\Phi(M)$ represents the r.h.s. of Eq. \eqref{eq:MB-convective-moments} at $M$-moment, such that $\Phi(0)=\rho$, $\Phi(1)=\rho u_{\alpha} = j_{\alpha}$ and $\Phi(2)=\rho \mathbb{P}_{\alpha \alpha}=\rho (c_{\textrm{sound}}^2 + u^2_{\alpha})$. The lattice speed of sound can be $c_{\textrm{sound}}=\sqrt{\theta}$. $\chi_{(M),\alpha}^*$ are the $M$-Lagrange multipliers for the each of the $M=0,1,2$ MB-moments respectively. $\chi_{(0),\alpha}^*$ is $\chi_{(0)}^*$ since the density is a scalar quantity. 
Because of constants are immaterial \cite{BoydVandenbergheBook} \cite{OttingerBook} and with $\chi_{(M),\alpha}=\textrm{\Large{\textit{e}}}(\chi_{(M),\alpha}^*)$, the solution for $f_i$, which becomes the $f_i^{\textrm{eq}}$, yields a thermal product form  
\begin{eqnarray} \label{eq:Complete-product-form}
f_i^{\textrm{eq}}= \prod_{\alpha=\{x,\textrm{y,z}\}}^{d} \Bigg( W_{c_{\alpha, i}} \Bigg( \prod_{M=0}^{M_{\textrm{max}}} \chi_{(M),\alpha}^{c_{\alpha, i}^M}  \Bigg)  \Bigg), 
\end{eqnarray} 
where $d$ stands for the dimension of the problem.  The Lagrange multipliers are found upon substituting \eqref{eq:Complete-product-form} into the  constraints \eqref{eq:MB-convective-moments} up to $M=M_{\textrm{max}}$. The result for D$d$Q$3^d$ can be formulated in the following form 
\begin{eqnarray} \nonumber
 	f_i^{\textrm{eq}} &=& \rho \prod_{\alpha=\{x,\textrm{y, z}\}}^{d} W_{c_{\alpha,i}} \Bigg( \frac{\mathbb{P}_{\alpha \alpha} -c_1^2}{\theta-c_1^2} \Bigg)
 	\Bigg( \frac{\sqrt{\rho^2 \mathbb{P}_{\alpha \alpha}^2 - j_{\alpha}^2 c_1^2}}{\rho \mathbb{P}_{\alpha \alpha} - j_{\alpha} c_1}  \Bigg)^{c_{\alpha,i}/c_1} \\ \label{eq:ExactThermalFlowEquationRaul-2D3D} 
 	&\times& \textcolor{green}{\underline{\textcolor{black}{  \Bigg( \frac{(\theta-c_1^2) \sqrt{\rho^2 \mathbb{P}_{\alpha \alpha}^2 - j_{\alpha}^2 c_1^2} }{\rho \theta (\mathbb{P}_{\alpha \alpha} - c_1^2)} \Bigg)^{c_{\alpha,i}^2/c_1^2} }}},
\end{eqnarray}
where the underlined term appears when $M_{\textrm{max}}=2$. 
Note that the insertion of $\theta=\theta_0=c_1^2/3$, $c_1=1$ and $j_{\alpha}=\rho u_{\alpha}$ into the one-dimensional weights \eqref{eq:ThermalWeightsD1Q3} with $\mu=0$ and also into Eq.  \eqref{eq:ExactThermalFlowEquationRaul-2D3D} reveals the reduced formulation (3) in \cite{AsinariKarlin2009}. It is interesting to note that no product form is mentioned in \cite{AsinariKarlin2009}. In addition,  the product form in \cite{ChikatamarlaKarlin2009} is only up to $M_{\textrm{max}}=1$ in Eq. \eqref{eq:Complete-product-form}, while no (thermal) product form is implemented in \cite{YudistiawanAnsumali2010} (c.f. Eq. (25) therein).  

When $M_{\textrm{max}}=1$ in \eqref{eq:Complete-product-form}, the result becomes similar as in \eqref{eq:ExactThermalFlowEquationRaul-2D3D} but without the underlined term and fixed $\theta=c_1^2/3$, $c_1=1$, meaning that only the density and the momentum density are fulfilled.   
Therefore, there exist spurious velocity terms in the (trace of the) pressure tensor $P_{\alpha \alpha}$ for the ELB method with $M_{\textrm{max}}=1$, as seen table \ref{tab:1DH2H3E} for $M=2$, denoted by $\textrm{E}_{(1)}^3$.  This lack of accuracy is solved by using $M_{\textrm{max}}=2$ in \eqref{eq:Complete-product-form} leading to  \eqref{eq:ExactThermalFlowEquationRaul-2D3D}, c.f. $\textrm{E}_{(2)}^3$ in table \ref{tab:1DH2H3E} for $M=2$.  However, because of the ELB  method is based on macroscopic descriptions, the value of $M_{\textrm{max}}$  is limited, e.g. up to 2,  and thus the increase of the lattice set $z>1$ cannot be equated with a similar increase in  $M_{\textrm{max}} $. This leads to spurious velocity terms in high-order ELB models, similar to those found in $\textrm{E}_{(1)}^3$ but for higher order (MB) moments. 
This, even after using $\theta$ as a helping parameter in an effort to match (MB) moments.  
The exactness is lost with the presence of spurious terms and thus the main LB idea is not fulfilled for high-order ELB models.

\bibliography{RaulMACHADO3PRE-PUBLICversion}%

\end{document}